\documentclass[12pt]{article}

\usepackage{latexsym,amsmath,amssymb,theorem,epsfig}

\usepackage{graphicx}
\usepackage{graphicx,subfigure}
\usepackage{epstopdf}
\usepackage[body={17.5cm, 21cm},right=2cm]{geometry}
\usepackage{amssymb}
\usepackage{amsmath}
\usepackage{psfrag}
\usepackage{epsfig}
\usepackage{cancel}
 \allowdisplaybreaks[4]

\usepackage[all]{xy}

\DeclareFontFamily{OT1}{rsfs10}{}
\DeclareFontShape{OT1}{rsfs10}{m}{n}{ <-> rsfs10 }{}
\DeclareMathAlphabet{\mathscript}{OT1}{rsfs10}{m}{n}

\numberwithin{equation}{section}

\newcommand{\ns}{\normalsize}

\newcommand{\be}{\beta}

\newcommand{\nn}{\cal{N}}

\def\gsim{ \lower .75ex \hbox{$\sim$} \llap{\raise .27ex \hbox{$>$}} }
\def\lsim{ \lower .75ex \hbox{$\sim$} \llap{\raise .27ex \hbox{$<$}} }
\def\be{\begin{equation}}
\def\ee{\end{equation}}
\def\bea{\begin{eqnarray}}
\def\eea{\end{eqnarray}}
\def\be{\begin{equation}}
\def\ee{\end{equation}}
\def\beq{\begin{eqnarray}}
\def\eeq{\end{eqnarray}}
\def\bse{\begin{subequations}}
\def\ese{\end{subequations}}
\def\nn{{\nonumber}}
\def\l{\left (}
\def\r{\right )}

\def\bo{{\square}}

\def\mn{{\mu\nu}}

\def\pd{{\dot{\pi}}}

\def\pdd{{\ddot{\pi}}}

\def\add{{\ddot{a}}}

\def\oh{{\frac{1}{2}}}

\begin{document}

\begin{titlepage}

\vspace{-5cm}

\title{
  \hfill{\ns }  \\[1em]
   {\LARGE Galileon Cosmology}
\\[1em] }
\author{
   Nathan Chow and Justin Khoury
     \\[0.5em]
   {\ns Department of Physics and Astronomy, University of Pennsylvania,}\\[-0.2cm]
   {\ns Philadelphia, PA 19104, USA}\\
{\ns Perimeter Institute for Theoretical Physics, Waterloo, Ontario N2L 2Y5, Canada}\\[0.3cm]}

\date{}

\maketitle

\begin{abstract}
We study the cosmology of a galileon scalar-tensor theory, obtained by covariantizing the decoupling lagrangian of the Dvali-Gabadadze-Poratti (DGP) model. Despite being local in 3+1 dimensions, the resulting cosmological evolution is remarkably similar to that of the full 4+1-dimensional DGP framework, both for the expansion history and the evolution of density perturbations. As in the DGP model, the covariant galileon theory yields two branches of solutions, depending on the sign of the galileon velocity. Perturbations are stable on one branch and ghost-like on the other. An interesting effect uncovered in our analysis is a cosmological version of the Vainshtein screening mechanism: at early times, the galileon dynamics are dominated by self-interaction terms, resulting in its energy density being suppressed compared to matter or radiation; once the matter density has redshifted sufficiently, the galileon becomes an important component of the energy density and contributes to dark energy. We estimate conservatively that the resulting expansion history is consistent with the observed late-time cosmology, provided that the scale of modification satisfies $r_c \;\gsim\; 15$~Gpc.
\end{abstract}
{\let\thefootnote\relax\footnotetext{{\tt lchow@sas.upenn.edu}}
\footnote{\tt jkhoury@sas.upenn.edu}}

\thispagestyle{empty}

\end{titlepage}

\section{Introduction}

Scalar-tensor theories of gravity have experienced a resurgence of sorts, over the last
twenty years. This is due in part to string theory, where the plethora of compactification moduli generically appear in the 4$D$ effective
theory with kinetic mixing with the graviton. Moreover, the discovery of accelerated expansion makes the
possibility that General Relativity is modified on the largest scales plausible. If this is the case, then the new gravitational degrees of freedom
relevant on cosmological scales are likely to include a scalar cousin for the graviton.

The best-known example of a scalar-tensor theory is due to Brans and Dicke (BD)~\cite{BD},
\be
S_{\rm BD} = \frac{M_{\rm Pl}^2}{2}\int {\rm d}^4x\sqrt{-g}\left(\Phi R - \frac{\omega_{\rm BD}}{\Phi}(\partial\Phi)^2\right) + \int {\rm d}^4x\sqrt{-g}{\cal L}_{\rm matter}[g] \,,
\label{bdintro}
\ee
where the matter Lagrangian is independent of $\Phi$. Unfortunately, the BD parameter is so tightly constrained by solar system and pulsar observations,
 $\omega_{\rm BD} \;\gsim\; 4\times 10^4$~\cite{BDcons}, that the cosmological effects of the BD scalar are rendered uninterestingly small. 

A tantalizing alternative is that the apparent decoupling of the scalar field is a local effect, owing to the large matter density of the solar system or pulsar
environment. In other words, the BD parameter is effectively a growing function of the density. While decoupled locally, the scalar field can have interesting
cosmological effects in the much sparser cosmic environment. There are only two robust mechanisms that realize this idea. One is the chameleon
mechanism~\cite{cham1,cham2,cham3}: by adding a suitable potential $V(\Phi)$, the scalar field acquires mass which depends on the density. The mass
is large in regions of high density, thereby suppressing any long-range interactions. Theories of $f(R)$ gravity~\cite{fR} rely on the chameleon effect to ensure consistency with solar system tests~\cite{wayne}.

A second mechanism is the Vainshtein screening effect of the longitudinal graviton or brane-bending mode, usually denoted by $\pi$, in the DGP model~\cite{DGP}. As we review in Sec.~\ref{decouple}, this effect is most easily understood in a certain decoupling limit of the theory~\cite{luty,nicolis}: $M_{\rm Pl}, M_5\rightarrow\infty$, keeping the strong coupling scale $(M_{\rm Pl}r_c^{-2})^{1/3}$ fixed. The resulting theory is local on the brane, and describes a self-interacting scalar field coupled to weak-field gravity in 3+1 dimensions:
\be
{\cal L} = -\frac{M_{\rm Pl}^2}{4}h^{\mu\nu}({\cal E}h)_{\mu\nu} + M_{\rm Pl}^2\pi\eta^{\mu\nu}({\cal E}h)_{\mu\nu} -
\frac{r_c^2}{M_{\rm Pl}}(\partial\pi)^2\square\pi + \frac{1}{2}h^{\mu\nu}T_{\mu\nu}\,,
\label{pilagintro}
\ee
where ${\cal E}_{\mu\nu}^{\;\;\alpha\beta} h_{\alpha\beta} = -\square h_{\mu\nu}/2 + \ldots$ is the linearized Einstein tensor. 
As a vestige of 5$D$ Lorentz transformations, the $\pi$ action is invariant under the Galilean shift symmetry, $\partial_\mu\pi \rightarrow \partial_\mu\pi + c_\mu$.
Thus $\pi$ has been dubbed a {\it galileon} field~\cite{galileon}. In regions of high density, $\rho\gg M_{\rm Pl}^2r_c^{-2}$, non-linearities in $\pi$ dominate and result in its decoupling. This is qualitatively similar to the chameleon mechanism, except that the galileon relies on derivative interactions as opposed to a scalar potential.
\newpage
In this paper, we study the cosmology of the galileon, by promoting~(\ref{pilagintro}) to a fully covariant, non-linear theory of gravity coupled to a galileon field:
\be
 S = \int {\rm d}^4x\sqrt{-g}\left(\frac{M_{\rm Pl}^2}{2}e^{-2\pi/M_{\rm Pl}}R - \frac{r_c^2}{M_{\rm Pl}}(\partial\pi)^2\square\pi + {\cal L}_{\rm matter}[g]\right)\,.
\label{nonlinintro}
\ee
As in the DGP model, where the Galilean shift symmetry is only exact in the strict decoupling limit $M_{\rm Pl}\rightarrow \infty$, the
shift symmetry is now broken by $M_{\rm Pl}$-suppressed operators in~(\ref{nonlinintro}). Of course the above non-linear completion is by no means unique --- many other Lagrangians, for instance including a $(\partial\pi)^4/M_{\rm Pl}^4$ term, will reduce to~(\ref{pilagintro}) in the weak-field limit. Since the cosmological predictions should be fairly robust under such corrections, however, we take~(\ref{nonlinintro}) as a fiducial galileon theory and study its implications for cosmology.

Despite being a local theory in 3+1 dimensions, the cosmology derived from~(\ref{nonlinintro}) comes remarkably close to reproducing that of the 4+1-dimensional
DGP model, at least for $Hr_c \;\gsim \; 1$. As we will see, the agreement holds for both the expansion history and the evolution of density perturbations. For a sneak preview, see Figs.~\ref{DGPcomp} and~\ref{DGPcomppert}, respectively. More generally, our galileon cosmology reproduces many qualitative features of DGP: 

\begin{enumerate}

\item The Friedmann equation allows for two branches of solutions, depending on the sign of $\dot{\pi}$. In analogy, the modified Friedmann equation in DGP~\cite{cedric} also has two branches:
\be
H^2 = \frac{\rho}{3M_{\rm Pl}^2} \pm \frac{H}{r_c}\,.
\label{DGPfriedintro}
\ee

\item One branch of solutions has stable perturbations, whereas the other is plagued with ghost-like instabilities. This again agrees with DGP~\cite{nicolis},
where the ``$-$" and ``$+$" branches in~(\ref{DGPfriedintro}) are stable and unstable, respectively.

\item The effective equation of state for the galileon satisfies $w_\pi < -1$ on the stable branch, and $w_\pi>-1$ on the unstable branch. This agrees
with the effective equation of state inferred from the $H/r_c$ correction in~(\ref{DGPfriedintro})~\cite{w<-1}.

\item Moreover, the two branches are classically disconnected, unless $R < 0$. This is closely related to the condition $\rho-3P \leq -12 M_{\rm Pl}^2/r_c^2$
necessary to transition from the stable to the unstable branch of solutions in the decoupling theory~\cite{nicolis}. 

\end{enumerate}

\noindent An important difference with DGP, however, is that our covariant galileon theory does not allow for self-accelerated cosmology --- the self-accelerated solution is spoiled by $1/M_{\rm Pl}$ terms in~(\ref{nonlinintro}). 

One of our key results is a cosmological analogue of the Vainshtein screening mechanism. At early times, $Hr_c\gg 1$, the dynamics of $\pi$ are dominated by the cubic interaction term, resulting in the galileon energy density being suppressed by ${\cal O}(1/Hr_c)$ compared to the matter or radiation fluid. When the matter density has dropped sufficiently, so that $Hr_c\sim 1$, the galileon becomes an important component of the total energy density and contributes to dark energy. 

We also study the effects of the galileon on the growth of inhomogeneities. By virtue of its non-minimal coupling to gravity, the galileon
enhances the gravitational attraction between particles, which translates into more efficient growth of density perturbations.
The screening mechanism is also at play in the evolution of perturbations: the galileon enhancement is suppressed for $Hr_c\gg 1$,
but becomes important once $Hr_c\sim 1$. A similar time-like Vainshtein effect was also observed in~\cite{degrav}.

While a full likelihood comparison to data is left for future study, we discuss various constraints on the galileon cosmology, such as
from estimates of the matter density at different redshifts, the luminosity distance relation, and the angular-diameter distance to the
last scattering surface. The resulting bound on $r_c$ is
\be
r_c\;\gsim \; 15\;{\rm Gpc}\,,
\label{rcboundintro}
\ee
which constrains the scale of the modification to be at least a few times the Hubble radius today.
However, since all other cosmological parameters are kept fixed in our considerations,~(\ref{rcboundintro}) is likely a conservative estimate. 

The paper is organized as follows. In Sec.~\ref{decouple} we review how the weak-field action~(\ref{pilagintro}) arises from the decoupling limit of the DGP model,
and describe the origin of the self-screening mechanism near spherical sources. In Sec.~\ref{NLextend} we discuss the non-linear
extension~(\ref{nonlinintro}) and derive the covariant equations of motion. We present in Sec.~\ref{sect:cosmo} the cosmology of the galileon model.
In particular, we derive an approximate analytic solution, which displays the screening mechanism, and study its stability. Our
analytic considerations are borne out by the numerical solutions presented in Sec.~\ref{sect:num}. Turning to inhomogeneities, we study in Sec.~\ref{pert}
the effects of the galileon on the growth of density perturbations. In Sec.~\ref{obssec} we discuss various observational constraints on galileon cosmology
and derive the bound on $r_c$ given in~(\ref{rcboundintro}). We conclude in Sec.~\ref{conclusions} with a brief summary and discuss future research
avenues. 

While preparing this manuscript we became aware that Deffayet {\it et al.} were independently studying a model with some similarities to ours~\cite{dpe,dpsv}.
This paper is based on the M.Sc. thesis of N.C. at the University of Waterloo~\cite{nathanthesis}. 

\section{Decoupling Limit of DGP}
\label{decouple}

In the DGP model, our visible universe is confined to a 3-brane in a 4+1-dimensional bulk. Despite the fact that the extra dimension is infinite in extent, 4$D$ gravity
is nevertheless recovered over some range of scales on the brane because of a 3+1-dimensional Einstein-Hilbert term, intrinsic to the brane:
\be
S_{\rm DGP} = \int_{\rm bulk} {\rm d}^5x\sqrt{-g_5}\frac{M_5^3}{2}R_5 +\int_{\rm brane} {\rm d}^4x \sqrt{-g_4} \left(\frac{M_{\rm Pl}^2}{2}R_4 + {\cal L}_{\rm matter}[g]\right)\,.
\ee
The bulk and brane Planck masses define a cross-over scale, 
%% suggestion 1
\be
r_c = \frac{M_{\rm Pl}^2}{2M_5^3}\,,
\ee
%% /suggestion 1
which separates the 4$D$ and 5$D$ regimes. At distances $r \ll r_c$ on the brane, the gravitational force law scales as $1/r^2$, whereas for $r\gg r_c$ it scales as $1/r^3$. 

From the point of view of a brane observer, the 5 helicity-2 states of the massless 5$D$ graviton combine to form a massive spin-2 representation in 4$D$. More precisely, the 4$D$ graviton is a resonance --- a continuum of massive states --- whose spectral width is peaked at the scale $r_c^{-1}$. As in massive gravity~\cite{vainshtein}, the helicity-0 or longitudinal mode, denoted by $\pi$, becomes strongly coupled at a much lower scale than $M_5$, given by~\cite{luty,ddgv}
\be
\Lambda_{\rm strong} = \left(M_{\rm Pl}r_c^{-2}\right)^{1/3}\,.
\label{Lamb}
\ee
For $r_c \sim H_0^{-1} = 10^{28}\;{\rm cm}$, for instance, this gives $\Lambda^{-1}_{\rm strong} \sim 1000$~km.
The strong coupling behavior is essential to the phenomenological viability of the model through the Vainshtein screening effect~\cite{vainshtein,ddgv}. As we review below, non-linear interactions in $\pi$ are important near an astrophysical source and result in the decoupling of $\pi$ from the source. The characteristic scale
below which $\pi$ is strongly coupled, denoted by $r_\star$, is given by
\be
r_\star = (r_c^2r_{\rm Sch})^{1/3}\,,
\label{r*}
\ee
where $r_{\rm Sch}$ is the Schwarzschild radius of the source. And since $r_c$ is cosmologically large (of order of the Hubble radius today), $r_\star$ is
parametrically larger than $r_{\rm Sch}$. 

In analogy with massive gravity~\cite{ags}, it is instructive to zoom in on the non-linearities in $\pi$ by considering the decoupling limit~\cite{luty,nicolis}:
$M_{\rm Pl},M_5\rightarrow\infty$ keeping the strong coupling scale $\Lambda_{\rm strong}$ fixed. Equivalently, around a spherical source this corresponds to sending $r_{\rm Sch}\rightarrow 0$ keeping $r_\star$ fixed. In other words, in this limit non-linearities in the helicity-2 (Einsteinian) modes drop out, while interactions of the helicity-0 state
survive. The resulting effective theory is local on the brane and describes (weak-field) gravity plus a scalar field $\pi$ in 3+1 dimensions:
\be
{\cal L}_{\rm Einstein} = -\frac{M_{\rm Pl}^2}{4}\tilde{h}^{\mu\nu}({\cal E}\tilde{h})_{\mu\nu} - 3(\partial\pi)^2 - \frac{r_c^2}{M_{\rm Pl}}(\partial\pi)^2\square\pi + \frac{1}{2}\tilde{h}^{\mu\nu}T_{\mu\nu} + \frac{1}{M_{\rm Pl}}\pi T\,,
\label{pilag}
\ee
where ${\cal E}_{\mu\nu}^{\;\;\alpha\beta}\tilde{h}_{\alpha\beta} = -\square \tilde{h}_{\mu\nu}/2 + \ldots$ is the linearized Einstein tensor. This lagrangian is, up to a total derivative term, invariant under the Galilean shift symmetry,
\be
\partial_\mu\pi \rightarrow \partial_\mu\pi + c_\mu\,,
\label{gal}
\ee
which is a vestige of the full 5$D$ Lorentz transformations. Thus $\pi$ has been dubbed a galileon field~\cite{galileon}.

\subsection{Self-screening effect}

The approximate recovery of general relativity in the vicinity of astrophysical sources, through the Vainshtein effect,
can be understood at the level of~(\ref{pilag})~\cite{nicolis}. The equation of motion for the galileon,
\be
\partial^\mu\left(6M_{\rm Pl}\partial_\mu\pi + 2r_c^2 \partial_\mu\pi\square\pi - r_c^2\partial_\mu(\partial\pi)^2\right) = - T \,,
\label{pieom}
\ee
is remarkable in many respects. Even though the interaction term in~(\ref{pilag}) contains four derivatives, the equation of motion is nevertheless
second-order --- all higher-derivative terms cancel out when performing the variation. Moreover,~(\ref{pieom}) takes the simple form $\partial_\mu j^{\;\mu}_\pi = -T/2M_{\rm Pl}$ for some $\pi$-current $j^{\;\mu}_\pi$, thereby allowing for a generalized ``Gauss' law": spherically-symmetric exterior solutions for $\pi$ only depend on the mass
enclosed. 

Let us indeed study the spherically-symmetric galileon profile due to a point mass: $T= -M\delta^3(r)$. In this case,~(\ref{pieom}) can be integrated to give
\be
6M_{\rm Pl} \pi'(r) + 4r_c^2\frac{\pi'^2(r)}{r} = \frac{M_{\rm Pl}^2r_{\rm Sch}}{r^2}\,.
\ee
Using~(\ref{r*}), the solution to this algebraic equation for $\pi'$ is given by
\be
\frac{\pi'(r)}{M_{\rm Pl}} = \frac{3r}{4r_c^2}\left(-1+\sqrt{1+\frac{4}{9}\frac{r_\star^3}{r^3}}\right)\,.
\ee
Note that we have chosen the branch of the solution such that $\pi'\rightarrow 0$ as $r\rightarrow\infty$. The other branch, corresponding to $\pi'$ diverging at infinity, belongs to the same branch of solutions as the self-accelerated DGP cosmology, and is therefore unstable~\cite{nicolis}. This is a general property: solutions to~(\ref{pieom}) always come in a pair, with one member continuously connected to the trivial solution, with stable perturbations, and the other connected to the self-accelerated cosmological solution, with unstable perturbations. It is impossible to classically move from one branch of solutions to the other without violating some energy condition~\cite{nicolis}. In this work, we focus almost exclusively on the stable branch of solutions.

At short distances, $r\ll r_\star$, the galileon-mediated force is clearly suppressed compared to the gravitational force:
\be
\frac{F_\pi}{F_{\rm grav}} = \frac{|\vec{\nabla}\pi|}{M_{\rm Pl}|\vec{\nabla}\Phi|} = \frac{r_\star^{3/2}}{r_c^2r^{1/2}}\frac{r^2}{r_{\rm Sch}} =  \left(\frac{r}{r_\star}\right)^{3/2}\ll 1\,.
\ee
Thus, as advocated, the strong interactions of $\pi$ lead to its decoupling near a source, and the theory reduces to Newtonian gravity.
This approximate recovery of standard gravity near a source has been established in approximate solutions of the full DGP model~\cite{ddgv,gruz,por}.
The above $\pi$-mediated force, albeit small in the solar system, is nevertheless constrained by lunar laser ranging observations~\cite{llr,moon,niayeshghazal}: $r_c\; \gsim\;  120$~Mpc. A comparable bound on $r_c$ has also been obtained by studying the effect on planetary orbits~\cite{battat}.

At large distances, $r \gg r_\star$, on the other hand, the non-linear terms in $\pi$ are negligible, and the resulting correction to Newtonian gravity is of order unity:
\be
\frac{F_\pi}{F_{\rm grav}} = \frac{1}{3}\,.
\label{weakreg}
\ee
The galileon-mediated force therefore leads to an enhancement of the gravitational attraction by a factor of 4/3. In this far-field regime, the theory reduces to a scalar-tensor theory, with the galileon acting as a Brans-Dicke scalar. 

\subsection{Jordan frame description}

Our action~(\ref{pilag}) is cast in Einstein frame, where the kinetic terms are diagonal, but $\pi$ couples directly to matter. We find it more convenient
to instead work in Jordan frame, by performing the shift
\be
h_{\mu\nu} = \tilde{h}_{\mu\nu} + \frac{2\pi}{M_{\rm Pl}}\eta_{\mu\nu}\,.
\ee
This removes the $\pi T$ coupling, at the price of introducing kinetic mixing between $h$ and $\pi$:
\be
{\cal L}_{\rm Jordan} = -\frac{M_{\rm Pl}^2}{4}h^{\mu\nu}({\cal E}h)_{\mu\nu} + M_{\rm Pl}^2\pi\eta^{\mu\nu}({\cal E}h)_{\mu\nu} - \frac{r_c^2}{M_{\rm Pl}}(\partial\pi)^2\square\pi + \frac{1}{2}h^{\mu\nu}T_{\mu\nu}\,.
\label{pilag2}
\ee
This forms makes the Brans-Dicke nature of the theory manifest, in the limit where the $\pi$-interactions can be neglected, with the Brans-Dicke parameter identified as $\omega_{\rm BD} = 0$.

\section{Non-linear Completion}
\label{NLextend}

Nearly all of the interesting phenomenological features of the DGP model are attributable to the helicity-0 mode $\pi$ and can be understood 
at the level of the decoupling theory. The Vainshtein effect, reviewed above, is one example. The existence of a self-accelerated solution is
another example: the equation of motion~(\ref{pieom}) in vacuum ($T = 0$) has a solution where $\pi \sim M_{\rm Pl}x_\mu x^\mu/r_c^2$, in agreement with
the weak-field limit of de Sitter space~\cite{nicolis}.

This motivates us to propose a 4$D$ theory of modified gravity, by promoting~(\ref{pilag2}) into a fully covariant, non-linear theory of gravity coupled to a
galileon field. By construction, this non-linear theory will reduce to~(\ref{pilag2}) in the limit of weak gravitational fields. Therefore its predictions will agree with
those of the full DGP model to leading order in $1/M_{\rm Pl}$. 

Looking at~(\ref{pilag2}), a natural non-linear completion suggests itself:
\be
 S = \int {\rm d}^4x\sqrt{-g}\left(\frac{M_{\rm Pl}^2}{2}e^{-2\pi/M_{\rm Pl}}R - \frac{r_c^2}{M_{\rm Pl}}(\partial\pi)^2\square\pi + {\cal L}_{\rm matter}[g]\right)\,,
\label{nonlin}
\ee
where $\square$ is now understood as the covariant Laplancian: $\square = \nabla^\mu\nabla_\mu$. This clearly reduces to~(\ref{pilag2}) in the weak-field limit.
The Galilean shift symmetry~(\ref{gal}) is softly broken in the action~(\ref{nonlin}) through $M_{\rm Pl}$-suppressed operators. This also true of the full DGP model,
where the Galilean symmetry arises only in the strict decoupling limit as a remnant of the full 5$D$ Lorentz group.

Of course the above non-linear extension is by no means unique. For instance, we could consider more general functions of $\pi/M_{\rm Pl}$ multiplying the Ricci scalar, or include corrections of the form $(\partial\pi)^4/M_{\rm Pl}^4$, all of which would drop out in the limit $M_{\rm Pl}\rightarrow\infty$. Be that as it may, we take~(\ref{nonlin}) as a fiducial covariant theory and explore its cosmological predictions. A study of more general lagrangians is left for the future.

Remarkably, as we will see in the next Section, the 4$D$ cosmology arising from~(\ref{nonlin}) reproduces many features of the full-fledged DGP model. Our Friedmann equation has two branches of solutions, depending on the sign of the velocity of $\pi$. The two branches are distinguished by having stable
or unstable (ghost-like) perturbations. Moreover, we uncover a cosmological analogue of the Vainshtein effect: at early times, when the density of the universe is high, non-linear interactions in $\pi$ are important, resulting in the galileon energy density being subdominant compared to the matter or radiation fluid. 

The covariant equation of motion for the galileon is readily obtained from~(\ref{nonlin}):
\be
\left(\square\pi\right)^2-\left(\nabla_\mu\nabla_\nu\pi\right)^2 - R^{\mu\nu}\nabla_\mu\pi\nabla_\nu\pi = \frac{M_{\rm Pl}^2}{2r_c^2}Re^{-2\pi/M_{\rm Pl}}\,.
\label{pieomcov}
\ee
Similarly, the Einstein equations are given by
\bea
\nonumber
e^{-2\pi/M_{\rm Pl}}M_{\rm Pl}^2 G_{\mu\nu}  &=& T_{\mu\nu} + M_{\rm Pl}^2\left(\nabla_\mu\nabla_\nu -g_{\mu\nu}\square\right) e^{-2\pi/M_{\rm Pl}} \\
&+& \frac{r_c^2}{M_{\rm Pl}}\left(2\nabla_\mu\pi\nabla_\nu\pi\square\pi + g_{\mu\nu}\nabla_\alpha\pi\nabla^\alpha(\partial\pi)^2- 2\nabla_{(\mu}\pi\nabla_{\nu)}(\partial\pi)^2\right)\,.
\label{einmod}
\eea
Since the matter action is independent of $\pi$, the matter stress-energy tensor satisfies the usual conservation law: $\nabla^\mu T_{\mu\nu} = 0$.

\section{Cosmology} 
\label{sect:cosmo}

In this Section we specialize the above equations to the cosmological context, by assuming homogeneity and isotropy. For simplicity, we focus on the case of a spatially-flat universe. Under these assumptions,~(\ref{pieomcov}) reduces to
\be
\label{JpiC}
\frac{{\rm d}}{{\rm d}t}\l H\pd^2\r + 3H\l H\pd^2\r = \frac{M_{\rm Pl}^2}{6r_c^2} Re^{-2\pi/M_{\rm Pl}}\,,
\ee
where dots represent derivatives with respect to proper time $t$. Remarkably, the left-hand side is reminiscent of the equation for a canonically-normalized
scalar field, $\ddot{\phi} + 3H\dot{\phi}$, with $\dot{\phi}$ identified as $H\dot{\pi}^2$. 

Because the effective field momentum is proportional to $\dot{\pi}^2$, however, the dynamics are quite different from those of a standard scalar field.
In particular, if $R\geq 0$, as is the case in a universe dominated by matter, radiation or vacuum energy, then solutions with $\dot{\pi}> 0$ and $\dot{\pi}< 0$
are classically disconnected. Indeed, expanding~(\ref{JpiC}),
\be
2H\dot{\pi}\ddot{\pi} + \ldots = \frac{M_{\rm Pl}^2}{6r_c^2} Re^{-2\pi/M_{\rm Pl}}\,,
\label{pottrans}
\ee
we see that $\dot{\pi}$ is driven away from zero if $R>0$. A necessary condition to transition from one branch to the other is therefore $R < 0$. This is closely related to the condition $\rho-3P \leq -12 M_{\rm Pl}^2/r_c^2$ necessary to transition from the stable to the unstable branch of solutions in the decoupling theory~\cite{nicolis}.
We will see that the above conclusions are borne out by numerical analysis --- as long as the universe is dominated by matter, radiation, or vacuum energy, $\dot{\pi}$ never changes sign.

%\begin{figure}[h!] %  figure placement: here, top, bottom, or page
%\begin{center}
%\includegraphics[width=0.9\textwidth]{pot.pdf}
%\end{center}
%   \caption{Treating the right-hand side of~(\ref{pottrans}) as an effective potential, we show $V\l\pi\r$ for generic $\pi$.} 
%   \label{potential}
%\end{figure}

We next turn our attention to the Friedmann and Raychaudhuri equations for the scale factor. For the matter, we assume as usual that the stress-energy tensor
is described by a perfect fluid,
\be
\label{pflu}
T_\mn = (\rho + P)u_\mu u_\nu + Pg_\mn\,,
\ee
with $\rho$ and $P$ denoting the energy density and pressure of the fluid. The Friedmann equation is given by the $(0,0)$ component of~(\ref{einmod}):
\beq
\label{JFriedrho}
3M_{\rm Pl}^2H^2e^{-2\pi/M_{\rm Pl}} = \rho + 6H\pd M_{\rm Pl}\l e^{-2\pi/M_{\rm Pl}} - \frac{r_c^2\pd^2}{M_{\rm Pl}^2}\r\,.
\eeq
Similarly, the Raychaudhuri equation follows as usual from a combination of the $(i,i)$ and $(0,0)$ components:
\bea
\nonumber
M_{\rm Pl}^2e^{-2\pi/M_{\rm Pl}}\frac{\add}{a} &=& -\frac{1}{6}(\rho+3P) -\frac{\pd^2}{2} \l \frac{3r_c^2\pdd}{M_{\rm Pl}} +4e^{-2\pi/M_{\rm Pl}} \r  \\
& +& \frac{1}{2} \l\frac{r_c^2\pd^2}{M_{\rm Pl}} + 2M_{\rm Pl} e^{-2\pi/M_{\rm Pl}}\r\l\pdd+2H\pd\r \,.
\label{JFriedp}
\eea

From the form of~(\ref{JFriedrho}) and~(\ref{JFriedp}), we can read off an effective energy density and pressure for the $\pi$ field, which we denote by $\rho_\pi$ and $P_\pi$. 
% JK: slight modif in the sentence below.
Of course, since $\pi$ is non-minimally coupled to gravity, $\rho_\pi$ is not conserved, hence it should only be understood as an effective energy density, informing us about the effects of the galileon on the cosmological evolution. In any event, we find:
\bea
\nonumber
\rho_\pi &=& 6H\pd M_{\rm Pl}\l e^{-2\pi/M_{\rm Pl}} - \frac{r_c^2\dot{\pi}^2}{M_{\rm Pl}^2}\r \;; \\
P_\pi &=& 2\dot{\pi}^2\left(\frac{r_c^2\ddot{\pi}}{M_{\rm Pl}}+2e^{-2\pi/M_{\rm Pl}}\right)-2M_{\rm Pl}e^{-2\pi/M_{\rm Pl}}\left(\ddot{\pi}+3H\dot{\pi}\right)\,.
\label{rhopi}
\eea
Note that $\rho_\pi$ changes sign depending on the choice of branch, {\it i.e.}, whether $\dot{\pi}$ is positive or negative.
That the galileon %% suggestion 2
effective %% /suggestion 2
energy density can be negative should not come as a surprise, since it is well-known that non-minimal couplings can
induce violations of various energy conditions in Jordan frame~\cite{DMDE1,DMDE2,staro}. This can also be seen
at the level of the effective equation of state,
\be
w_\pi  = \frac{P_\pi}{\rho_\pi} = \frac{-\ddot{\pi}\left(1-\frac{r_c^2\dot{\pi}^2}{M_{\rm Pl}^2}e^{2\pi/M_{\rm Pl}}\right) + \dot{\pi}\left(2\frac{\dot{\pi}}{M_{\rm Pl}}-3H\right)}
{3H\dot{\pi}\left(1-\frac{r_c^2\dot{\pi}^2}{M_{\rm Pl}^2}e^{2\pi/M_{\rm Pl}}\right)}\,,
\label{wpi}
\ee
which, {\it a priori}, allows for $w_\pi < -1$. In fact, we will see in Sec.~\ref{t*} that $w_\pi < -1$ holds at early times on the stable branch, when the universe is radiation- or matter-dominated. %% suggestion 3
We will also see in Sec.~\ref{results} that $w_\pi$ anyway contributes positively to the total effective equation of state, $w_{\rm tot}$ (see Fig.~\ref{wtots}), since $\rho_\pi$ is negative.
%% /suggestion 3

Many of these features also arise in the full DGP model. Indeed, if we think of the $H/r_c$ modification in the DGP Friedmann equation~\cite{cedric},
\be
H^2 = \frac{\rho}{3M_{\rm Pl}^2} \pm \frac{H}{r_c}\,,
\label{DGPfried}
\ee
as an effective contribution to the matter content, then clearly its energy density can have either sign, depending on the choice of branch. By the same token,
the effective equation of state corresponding to the modification can be $<-1$. For instance, on the normal branch, the effective energy density is negative
and $w_{\rm eff} <-1$~\cite{w<-1}. While such phantom behavior may at first seem surprising, our analysis now makes it clear that it is nothing but a natural consequence
of the scalar-tensor nature of gravity on the brane.

\subsection{Self-accelerated solution?}
\label{selfac}

In analogy with the full-fledged DGP model, we are tempted to look for a self-accelerated cosmology --- a de Sitter solution in the absence of any cosmological constant or
matter other than $\pi$ itself. In the regime $|\pi|\ll M_{\rm Pl}$ where we expect agreement with DGP, the equations of motion at first sight do seem to allow for approximate self-acceleration. Setting $\dot{\pi} = \dot{\pi}_0$ and $H=H_0$ to be constant in~(\ref{JpiC}) and~(\ref{JFriedrho}), one finds an approximate solution for $|\pi|\ll M_{\rm Pl}$, given by
\bea
\nonumber
\pi_0 &\approx & \sqrt{\frac{2}{3}}\frac{M_{\rm Pl}t}{r_c}\;; \\
H_0 &\approx & \left(\frac{2}{3}\right)^{3/2}\frac{1}{r_c}\,,
\eea
where the ``$\approx$"'s indicate the assumption $|\pi| \ll M_{\rm Pl}$. Up to a trivial redefinition of $r_c$, this agrees with the self-accelerated DGP solution that follows from the ``$+$" branch of~(\ref{DGPfried}): $H_0^{\rm DGP} = 1/r_c$.

Unfortunately, the approximation $|\pi|\ll M_{\rm Pl}$ breaks down within a time $t\sim r_c$, which is of the order of a Hubble time. In other words, $H$
evolves significantly over a Hubble time, and hence cannot be approximated as constant. Self-accelerated cosmology is spoiled in our galileon theory by
$\pi/M_{\rm Pl}$ corrections. 

\subsection{Early-time solution and cosmological screening}
\label{t*}

In this Section we derive approximate analytic solutions for when the universe is dominated by other components than $\pi$, such as matter, radiation or dark energy.
We will see that the dynamics of the galileon exhibit a time-like analogue of the Vainshtein effect: at early times, $t\ll r_c$, non-linearities in $\pi$ are
important, resulting in the galileon energy density being negligible. Once $t\sim r_c$, however, $\pi$ exits the strongly-coupled regime, and $\rho_\pi$
becomes a significant contribution to the total energy density.

To simplify the analysis, suppose that the universe is dominated by a single matter component with constant equation of state $w$. Moreover, we assume that the variation in $\pi$ throughout this phase is small in Planck units: $|\Delta\pi|\ll M_{\rm Pl}$. Since we can always set $\pi(t=0) = 0$ by trivial rescaling of $M_{\rm Pl}$, it follows that $e^{2\pi/M_{\rm Pl}}\approx 1$. The consistency of these approximations will be checked {\it a posteriori}.

With these assumptions, the Friedmann equation~(\ref{JFriedrho}) reduces to its standard form, $3H^2M_{\rm Pl}^2 \approx \rho$, with the usual solution
\be
a(t) \approx t^{\frac{2}{3(1+w)}}\,.
\ee
Substituting this into~(\ref{JpiC}),
\be
\frac{2}{H}\dot{\pi}\ddot{\pi} + \frac{3}{2}(1-w)\dot{\pi}^2 \approx \frac{M_{\rm Pl}^2}{2r_c^2}(1-3w)\,,
\ee
we see that  the galileon equation of motion allows for a solution with $\dot{\pi} = {\rm constant}$:
\be
\frac{\dot{\pi}}{M_{\rm Pl}} = \pm\frac{1}{r_c}\sqrt{\frac{1-3w}{3(1-w)}}\,.
\label{constpi}
\ee
Therefore a real solution exists if either $w\leq1/3$ or $w>1$, but, as we will see shortly, only for $w\leq 1/3$ is the solution a dynamical attractor.
Moreover, we will see in Sec.~\ref{stab} that among the two branches of solutions in~(\ref{constpi}), only $\dot{\pi} < 0$ has stable fluctuations.
The other branch, with $\dot{\pi}> 0$, is plagued with ghost-like instabilities.

\subsubsection{Cosmological screening}
\label{consistency}

The above constant-$\dot{\pi}$ solution is only valid provided that the galileon energy density is a negligible
contribution to the total energy. With our approximation $e^{2\pi/M_{\rm Pl}}\approx 1$, the first of~(\ref{rhopi}) reduces to
\bea
\nonumber
\frac{\rho_\pi}{3H^2M_{\rm Pl}^2} &=& \frac{2}{H}\frac{\dot{\pi}}{M_{\rm Pl}}\left(1-\frac{r_c^2\dot{\pi}^2}{M_{\rm Pl}^2}\right)\\
& =& \pm \frac{4}{Hr_c} \sqrt{\frac{1-3w}{3(1-w)}}\frac{1}{3(1-w)}\sim \frac{1}{Hr_c}\,.
\label{rhosub}
\eea
This elucidates the time-like screening effect advocated earlier: at early times, when $Hr_c\gg 1$ (or, equivalently, $t\ll r_c$), non-linear galileon interactions are important, and as a result its gravitational backreaction is negligible. In particular, this ensures that nucleosynthesis and recombination proceed as in standard cosmology, with negligible corrections coming from $\pi$. We note in passing that the parametric dependence of $\rho_\pi/3H^2M_{\rm Pl}^2$ is consistent with the modified Friedmann equation in the full DGP model: looking back at~(\ref{DGPfried}), the relative contribution of the $H/r_c$ modification term to the total expansion rate is indeed suppressed by $1/Hr_c$.

The self-screening effect breaks down after a time of order $r_c$, at which point $\rho_\pi$ becomes a significant
contribution to the expansion rate. As we will see in Sec.~\ref{obssec}, constraints on the late-time expansion history will enforce a lower bound on $r_c$.
Incidentally, $t\sim r_c$ also signals the moment when the approximation $|\pi|\ll M_{\rm Pl}$ breaks down, since 
\be
\left\vert\frac{\pi}{M_{\rm Pl}}\right\vert = \sqrt{\frac{1-3w}{3(1-w)}} \frac{t}{r_c}\,.
\ee
To summarize, the constant-$\dot{\pi}$ solution in~(\ref{constpi}) is valid at early times, $t\ll r_c$, when the galileon is strongly coupled. In this regime, the
galileon energy density can be consistently neglected, and the galileon excursion in field space is small in Planck units.

An illustrative way to understand this cosmological screening is to consider the non-linear galileon interactions as an effective BD parameter. By comparing the BD and galileon actions, given by~(\ref{bdintro}) and~(\ref{nonlin}) respectively, and making the identification $\Phi = e^{-2\pi/M_{\rm Pl}}$, we can define an effective dynamical $\omega_{\rm BD}^{\rm eff}$
\beq
\omega_{\rm BD}^{\rm eff} \l t \r &=& \frac{r_c^2}{2 M_{\rm Pl}}e^{2\pi/M_{\rm Pl}}\bo\pi \nn \\ 
&\approx& \mp\oh\sqrt{\frac{3\l1-3w\r}{1-w}}Hr_c\,,
\label{dynwbd}
\eeq
where in the second line we have made the approximation $e^{2\pi/M_{\rm Pl}} \approx 1$, and substituted the constant-$\dot{\pi}$ solution~(\ref{constpi}). Following the discussion of the previous paragraph, $\omega_{\rm BD}^{\rm eff}$ is large at early times, when $Hr_c\gg 1$, and standard cosmology is recovered. When $ Hr_c \sim 1$, however, we enter the scalar-tensor gravity regime with $\omega_{\rm BD}^{\rm eff} \sim 1$.

For completeness, we also derive the galileon equation of state during the strong coupling phase. Substituting~(\ref{constpi}) into~(\ref{wpi}), we obtain
\be
w_\pi  = -\frac{3}{2}(1-w) + {\cal O}\left(\frac{1}{Hr_c}\right)\,.
\ee
Interestingly, in this regime $w_\pi$ is completely fixed by the background equation of state. In particular, this says that $w_\pi \leq -1$ for $w\leq 1/3$. As discussed earlier,
however, this does not signal the presence of ghost instabilities, but instead is a natural consequence of the non-minimal coupling to gravity.

\subsubsection{Dynamical Attractor}
\label{dynact}

We next prove that our constant-$\dot{\pi}$ solution is a dynamical attractor for physically-relevant values of $w$. Perturb the galileon as 
\be
\pi(t) = \bar{\pi}(t) + \varphi(t)\,, 
\ee
with $\dot{\bar{\pi}}$ given by~(\ref{constpi}). Since the backreaction of $\rho_\pi$ on the expansion has already been shown negligible in this strongly coupled regime, $H(t)$ and $R(t)$ are oblivious to $\varphi$ and hence can be left unperturbed. Expanding~(\ref{JpiC}) to linear order in $\varphi$, while remembering that $e^{2\pi/M_{\rm Pl}}\approx 1$ in our approximation, we get
\be
\frac{{\rm d}}{{\rm d}t}\left(H\dot{\varphi}\right) + 3H\left(H\dot{\varphi}\right) \approx 0\,.
\ee
It follows that $H\dot{\varphi}\sim 1/a^3$, or
\be
\dot{\varphi} \sim \frac{1}{Ha^3}\sim t^{-\frac{1-w}{1+w}}\,.
\ee
Thus perturbations redshift away compared to $\dot{\bar{\pi}} = {\rm constant}$ for $-1 < w < 1$, and hence the solution is an attractor in this range.
On the other hand, from~(\ref{constpi}) we know that the constant-$\dot{\pi}$ solution only exists for either $w\leq 1/3$ or $w>1$. The above analysis
has now established that the solution is stable for $w\leq 1/3$ and unstable for $w>1$.

\subsubsection{General stability analysis}
\label{stab}

We can easily extend the above stability analysis to general perturbations: 
\be
\pi(\vec{x},t) = \bar{\pi}(t) + \varphi(\vec{x},t)\,.
\ee
We work at the level of the action, since here it is transparent whether perturbations have a right-sign kinetic term or are ghost-like --- a diagnosis that is trickier to make with linearized equations of motion.

Since $\pi$ is strongly coupled at early times, its perturbations are dominated by the cubic interaction term in~(\ref{nonlin}):
\be
 S_{\rm eff} =  -\frac{r_c^2}{M_{\rm Pl}} \int {\rm d}^4x\sqrt{-g} (\partial\pi)^2\square\pi\,.
\label{Seff}
\ee
Once again, as in the analysis of Sec.~\ref{dynact}, we neglect perturbations in the metric, since the cosmological background is driven by
some other source of stress-energy. Expanding~(\ref{Seff}) to quadratic order in $\varphi$, then after some integration by parts we obtain
\be
S_{\rm eff} ^{(2)} = -\frac{2r_c^2}{M_{\rm Pl}}\int {\rm d}^4x\sqrt{-g}\left(\square\bar{\pi} g_{\mu\nu} -\nabla_\mu\nabla_\nu\bar{\pi}\right)\partial^\mu\varphi\partial^\nu\varphi\,.
\label{Squad1}
\ee
Note that this holds for arbitrary background $\bar{\pi}(\vec{x},t)$. To make contact with the decoupling results of~\cite{nicolis}, it is useful to define $\bar{K}_{\mu\nu} = -r_c^2\nabla_\mu\nabla_\nu\bar{\pi}/M_{\rm Pl}$, which measures the extrinsic curvature of the brane in the decoupling limit of the full DGP model. In terms of $\bar{K}_{\mu\nu}$,~(\ref{Squad1}) takes the form
\be
S_{\rm eff} ^{(2)} = -2\int{\rm d}^4x \sqrt{-g}\left(\bar{K}_{\mu\nu} - g_{\mu\nu}\bar{K}\right)\partial^\mu\varphi\partial^\nu\varphi\,.
\label{SK}
\ee
In the weak-field limit, where we expect consistency with the decoupling limit, this indeed agrees with Eq.~(22) of~\cite{nicolis} in the strong coupling regime: $\bar{K}_{\mu\nu}\gg 1$. Remarkably, the form of the moduli space metric, $\bar{K}_{\mu\nu} - g_{\mu\nu}\bar{K}$, is preserved in our 4$D$ covariant theory.

Specializing to the constant-$\dot{\pi}$ solution,~(\ref{Squad1}) reduces to 
\be
S_{\rm eff} ^{(2)} = \frac{6r_c^2}{M_{\rm Pl}}\int{\rm d}^4x \sqrt{-g} H\dot{\bar{\pi}}\left(-\dot{\varphi}^2+\frac{2}{3}(\vec{\nabla}\varphi)^2\right)\,.
\ee
Thus the kinetic term of fluctuations is proportional to $\dot{\bar{\pi}}$ --- the kinetic term is positive if $\dot{\bar{\pi}} < 0$, corresponding to stable perturbations, and is negative if $\dot{\bar{\pi}}> 0$, corresponding to ghost-like perturbations. Hence, just like in the full DGP model, one branch of solutions
is stable, whereas the other is unstable. Moreover, looking back at~(\ref{rhosub}), the stable branch has $\rho_\pi < 0$, whereas the unstable branch has $\rho_\pi > 0$. Again this is consistent with DGP --- interpreting the $H/r_c$ modification in the DGP Friedmann equation~(\ref{DGPfried}) as an effective energy density,
then this energy density is negative on the stable (minus-sign) branch and positive on the unstable (positive-sign) branch.

\section{Numerical Analysis}
\label{sect:num}

We can solve the $\pi$ equation of motion~(\ref{JpiC}) and the Raychaudhuri equation~(\ref{JFriedp}) numerically to obtain exact cosmological solutions.
We focus exclusively on the {\it stable branch}, by setting initial conditions with $\dot{\pi} < 0$. Since $\rho_\pi < 0$ on this branch, we must include a dark energy
component to \\ 

\begin{figure}[h!] %  figure placement: here, top, bottom, or page
\begin{center}
\includegraphics[width=0.9\textwidth]{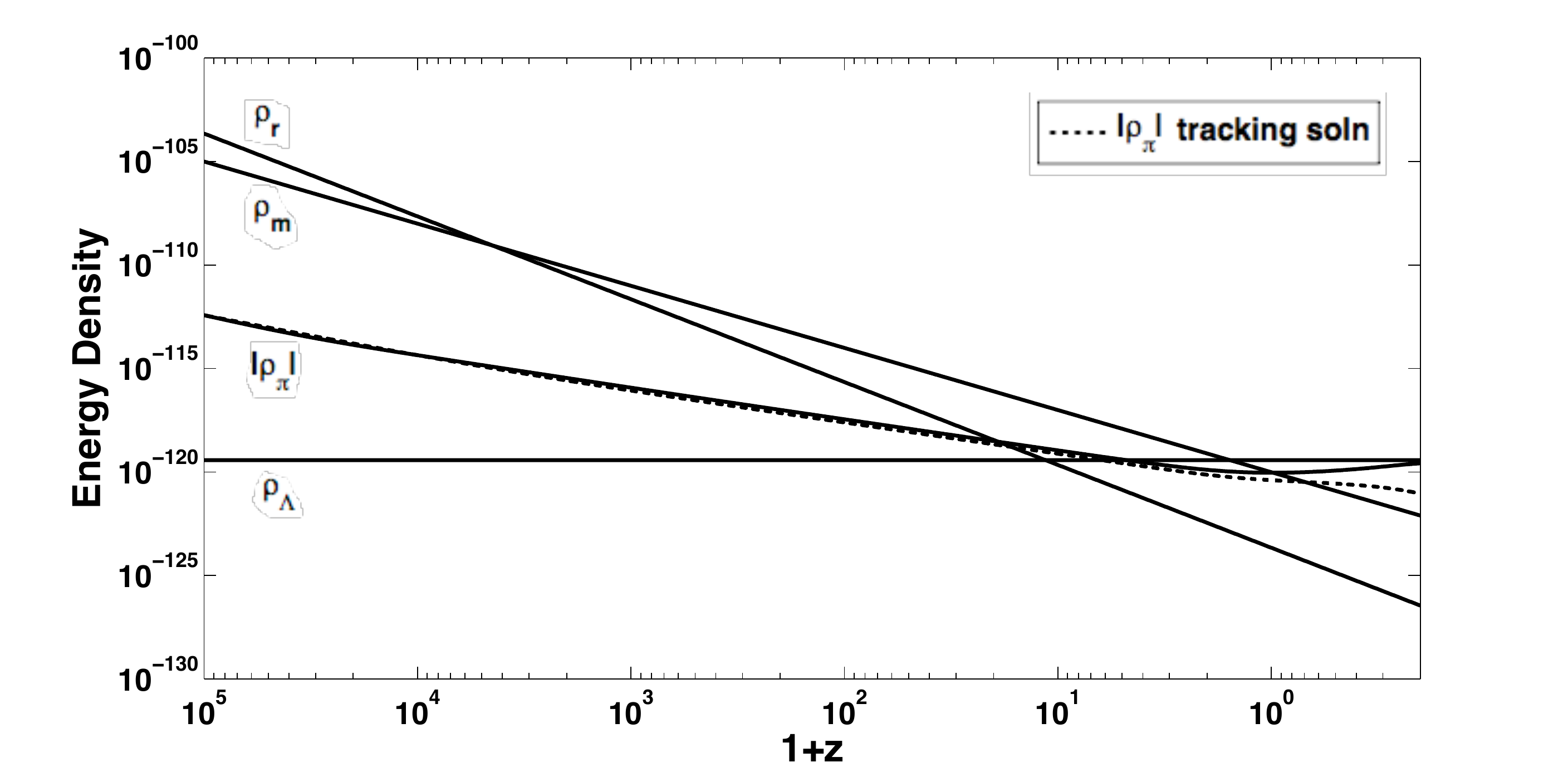}
\end{center}
   \caption{Result of numerically integrating the cosmological evolution equations with $r_c = 10$~Gpc. The solid curves denote the matter,
   radiation, cosmological constant and galileon energy density. The dotted line is the galileon energy density as predicted from the
   analytic solution.} 
   \label{rhos10}
\end{figure}

 \begin{figure}[h!] %  figure placement: here, top, bottom, or page
\begin{center}
\includegraphics[width=0.9\textwidth]{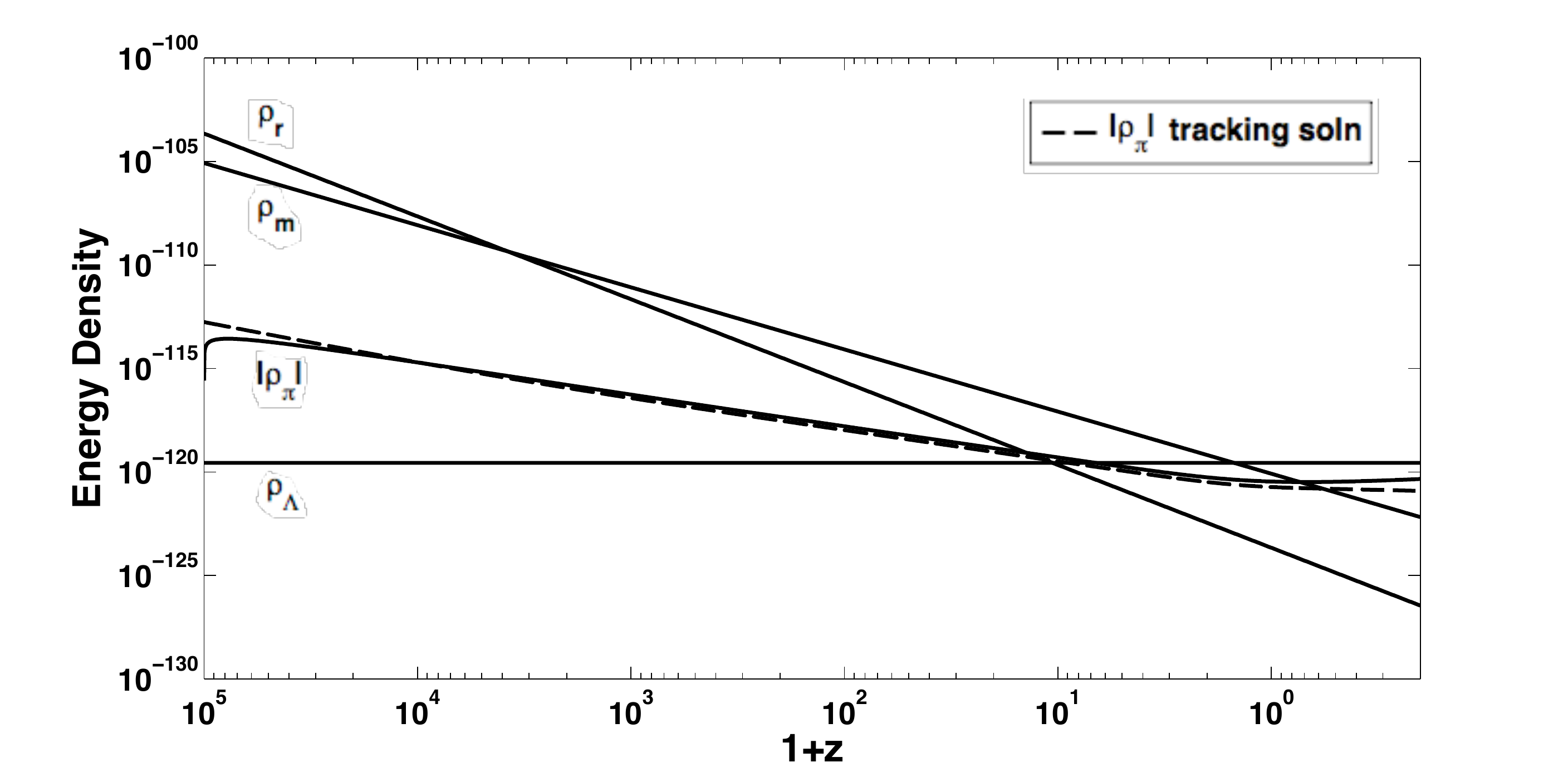}
\end{center}
   \caption{Same as Fig.~\ref{rhos10}, this time for $r_c = 20$~Gpc.} 
   \label{rhos20}
\end{figure}

\noindent obtain late-time acceleration, which we take to be a cosmological constant $\Lambda$ for simplicity. In other words, in addition to $\pi$ we include
radiation, matter and a cosmological term. And since $\rho_\pi$ contributes to the effective dark energy component, for each value for $r_c$ we adjust
$\Lambda$ so that the fractional contribution in matter today is kept fixed to the fiducial value $\Omega_{\rm m}^{(0)} = 0.26$.

\subsection{Results}
\label{results}

Figures~\ref{rhos10} and~\ref{rhos20} show the evolution of the energy density for the various components as a function of redshift, 
for $r_c = 10$ and $20$~Gpc, respectively. (Since these are log-log plots and $\rho_\pi < 0$, for the galileon component we instead plot $\left\vert\rho_\pi\right\vert$.)
These figures confirm the analytic results derived in Sec.~\ref{t*}. The dotted and dashed lines in Figs.~\ref{rhos10} and~\ref{rhos20}, respectively,
trace the galileon energy density as predicted by the constant-$\dot{\pi}$ solution of Sec.~\ref{t*}. Note that, to plot the analytic prediction, we have
substituted in~(\ref{constpi}) the total equation of state $w_{\rm tot}$ defined by 
\be
\frac{\dot{H}}{H^2} = -\frac{3}{2}(1+w_{\rm tot})\,. 
\label{wtotdef}
\ee

From the figures, we see that for most of the cosmological evolution, the actual galileon energy density agrees well with the constant-$\dot{\pi}$ prediction. Moreover, the constant-$\dot{\pi}$ solution is manifestly an attractor: as seen in Fig.~\ref{rhos20}, for instance, the galileon energy density quickly converges to the analytic prediction. The exact solution starts to deviate from the analytic prediction around the present time, however, when the Hubble radius $H^{-1}$ becomes comparable to $r_c$ --- this is consistent with the discussion of Sec.~\ref{consistency}. Note that the galileon energy density is a more significant contribution to the dark energy today for $r_c = 10$~Gpc than for $20$~Gpc.

\begin{figure}[h!] %  figure placement: here, top, bottom, or page
\begin{center}
\includegraphics[width=0.8\textwidth]{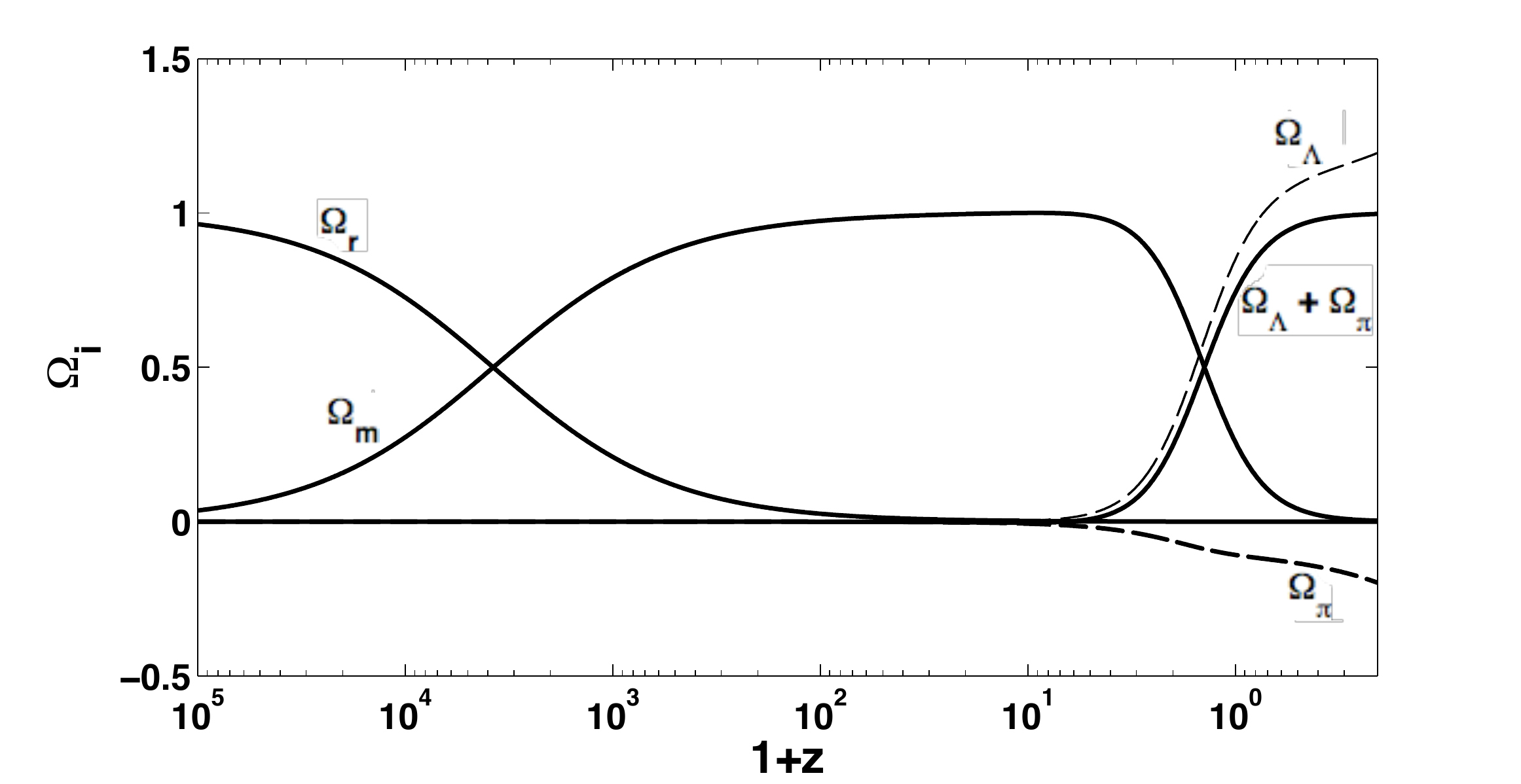}
\end{center}
   \caption{The fractional contributions to the total energy density from matter ($\Omega_{\rm m}$), radiation ($\Omega_{\rm r}$), cosmological constant ($\Omega_{\Lambda}$, dashed line) and the galileon field ($\Omega_{\pi}$, bold dashed line) for $r_c = 20$~Gpc, as a function of redshift.} 
   \label{Oms}
\end{figure}

Figure~\ref{Oms} shows the fractional contributions to the total energy density, defined for the various components as
\be
\Omega_i = \frac{\rho_ie^{2\pi/M_{\rm Pl}}}{3H^2M_{\rm Pl}^2}\,.
\label{Omi}
\ee 
Note that the galileon contribution is negative, since $\dot{\pi} < 0$. This figure shows once again that the galileon is subdominant
at early times, until $Hr_c \sim 1$ when $\rho_\pi$ becomes a significant contribution to the
expansion rate. 

 \begin{figure}[h!] %  figure placement: here, top, bottom, or page
\begin{center}
\includegraphics[width=0.8\textwidth]{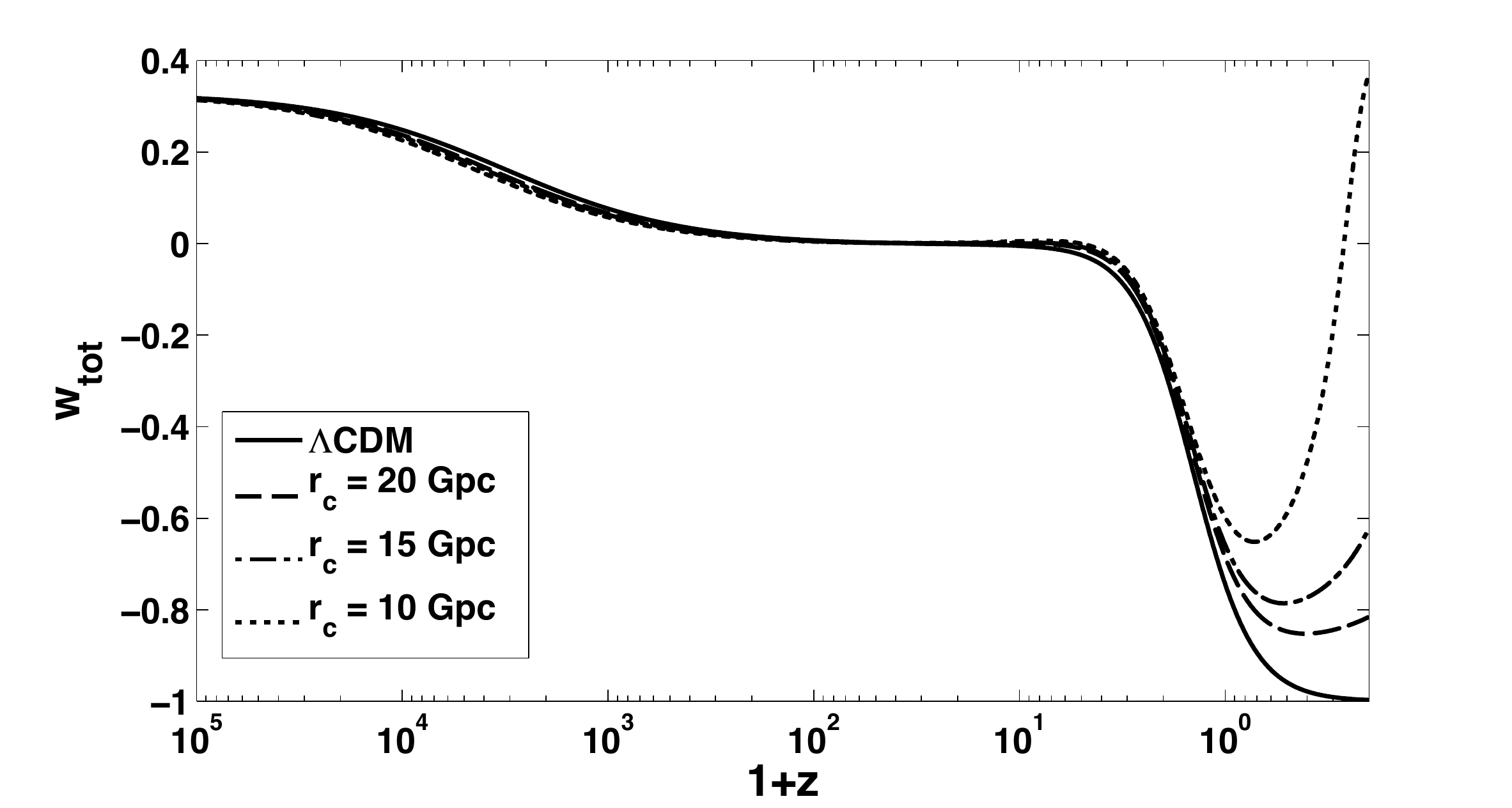}
\end{center}
   \caption{The effective equation of state for the expansion history, defined in terms of the Hubble parameter by $\dot{H}/H^2 = -3(1+w_{\rm tot})/2$, for $\Lambda$CDM (solid), and the galileon model with $r_c = 10$~Gpc (dotted), 15~Gpc (dash-dotted) and 20~Gpc (dashed).} 
   \label{wtots}
\end{figure}

\begin{figure}[h!] %  figure placement: here, top, bottom, or page
\begin{center}
\includegraphics[width=0.8\textwidth]{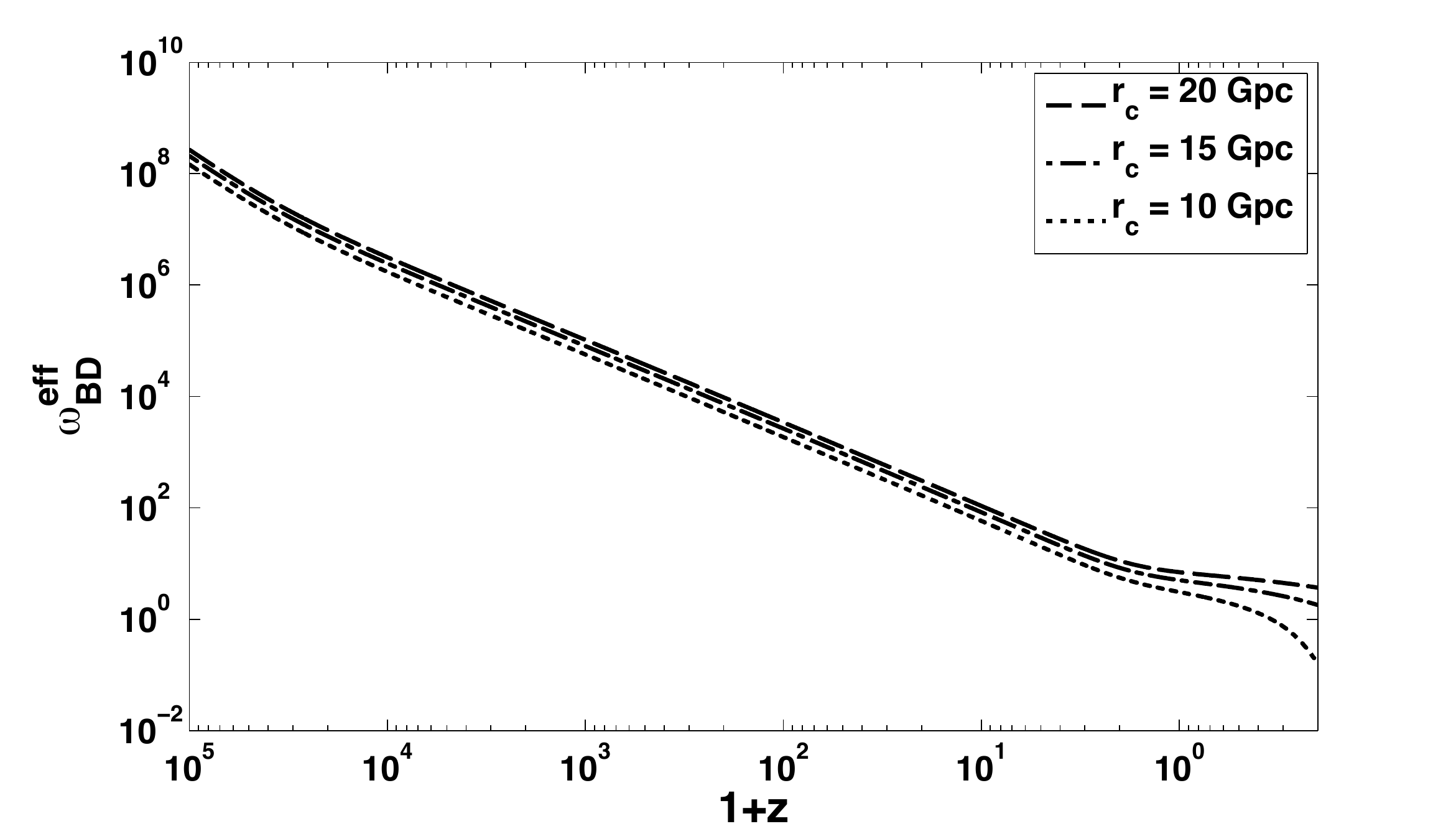}
\end{center}
   \caption{The effective BD parameter, $\omega_{\rm BD}^{\rm eff}$, with $r_c = 10$~Gpc (dotted), 15~Gpc (dash-dotted) and 20~Gpc (dashed).} 
   \label{wbds}
\end{figure}

Plotted in Fig.~\ref{wtots} is the effective equation of state for the whole evolution, $w_{\rm tot}$, defined in~(\ref{wtotdef}).
Here we plot the $\Lambda$CDM behavior (solid curve) compared to our galileon model with $r_c = 10$~Gpc (dotted curve), 15~Gpc (dash-dotted) and
20~Gpc (dashed). The early-time behavior is as expected from standard cosmology --- the equation of state goes through successive stages of
$w_{\rm tot}\approx 1/3$ and $w_{\rm tot}\approx 0$, corresponding to radiation- and
matter-dominated eras, respectively. At late times, the energy density is dominated by an effective dark energy component
composed of $\Lambda$ and $\rho_\pi$, with negative equation of state. Compared to $\Lambda$CDM, the galileon contribution
pushes the dark energy equation of state to values larger than $-1$. And the smaller $r_c$ is, the further $w_{\rm tot}$ is from $-1$ at late times.
Conversely, $w_{\rm tot}$ approaches $-1$ today as $r_c\rightarrow\infty$. While it may seem surprising at first sight that $w_{\rm tot} > -1$ at late times, since we argued in Sec.~\ref{sect:cosmo} that $w_\pi < -1$ on the branch of interest, one must keep in mind that $\rho_\pi < 0$ --- the galileon contribution to $\dot{H}$, being proportional to $-(1+w_\pi)\rho_\pi$, is therefore {\it negative}. \\
\indent To say a few words about the asymptotic behavior, since $\rho_\pi$ is negative and decreasing, it eventually catches up with $\rho_\Lambda$. At this point the Hubble parameter vanishes, and the universe starts to contract. Galileon models thus generically predict a late-time contracting phase. This could naturally match onto an ekpyrotic~\cite{ek1,ek2,ek3,ek4,ek5}, New Ekpyrotic~\cite{newek1,newek2,newek3,newek4,newek5} or cyclic~\cite{cyclic1,cyclic2} contracting phase, for instance. See~\cite{jlreview} for a review of these models. \\
\indent Figure~\ref{wbds} shows the effective BD parameter $\omega_{\rm BD}^{\rm eff}$, introduced in~(\ref{dynwbd}), for the whole evolution, with $r_c = 10$~Gpc (dotted curve), 15~Gpc (dash-dotted) and 20~Gpc (dashed). At early times the $\omega_{\rm BD}^{\rm eff} \sim 10^8$ is large, and standard cosmology is recovered. At late times, $\omega_{\rm BD}^{\rm eff} \sim 1$, and the theory reduces to scalar-tensor gravity. As expected, the scalar-tensor regime is achieved at earlier times for smaller values of $r_c$.

\subsection{Comparison with DGP cosmology}
\label{DGPc}

\begin{figure}[h!] %  figure placement: here, top, bottom, or page
\begin{center}
\includegraphics[width=0.9\textwidth]{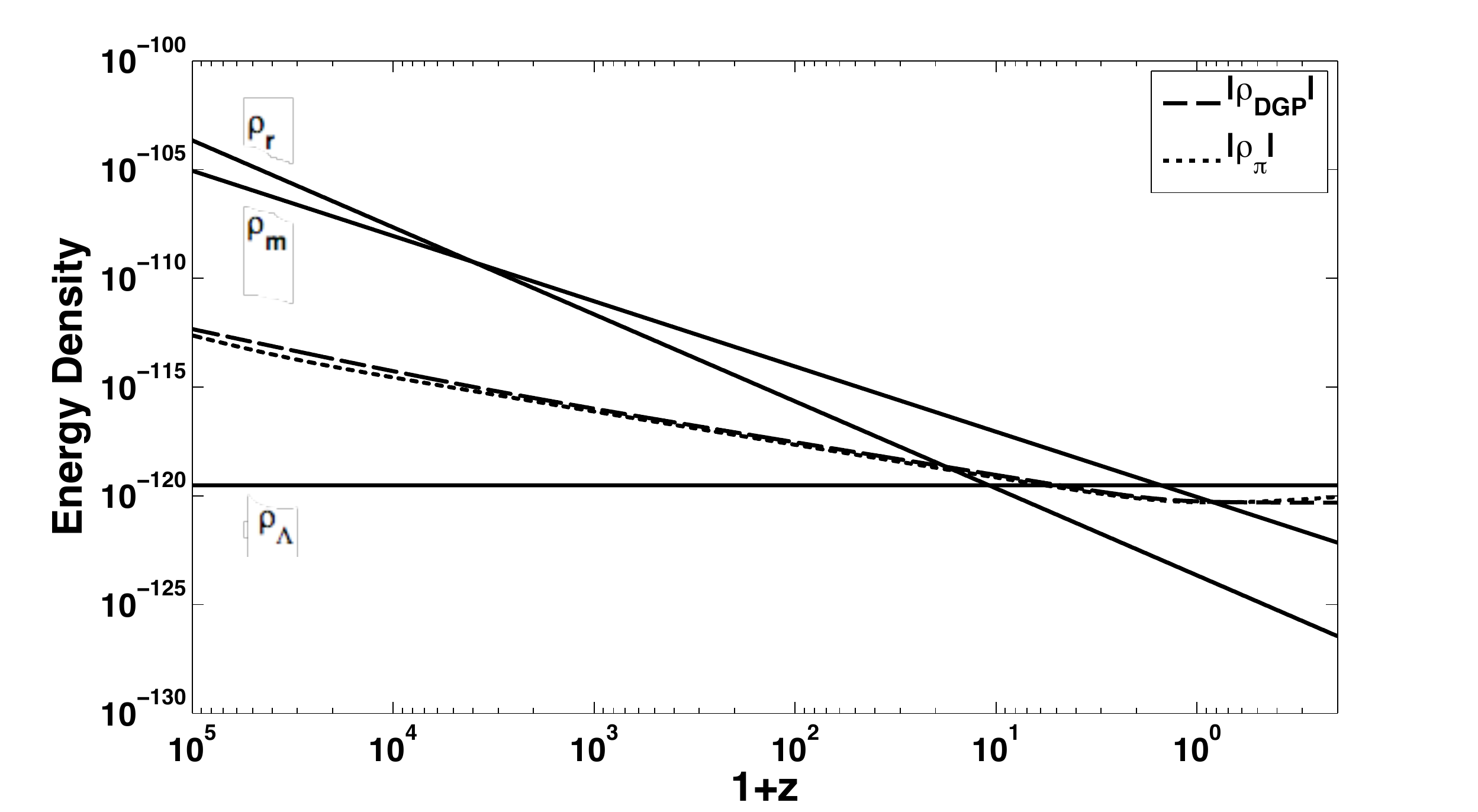}
\end{center}
   \caption{Comparison of the galileon energy density ($|\rho_\pi|$) with the effective energy density coming from the modification to the Friedmann equation in the DGP model ($|\rho_{\rm DGP}|$). We have used $r_c = 15$~Gpc in each case and kept the matter, radiation and cosmological constant densities fixed. The densities $\rho_\pi$ and $\rho_{\rm DGP}$ agree well for the entire expansion history until today, and begin to diverge in the future when $\pi\sim M_{\rm Pl}$.} 
   \label{DGPcomp}
\end{figure}

How closely does our 4$D$ galileon theory come to reproducing the cosmology of the 5D DGP model? To make the comparison it is convenient to
think of the $H/r_c$ correction term in the DGP Friedmann equation~(\ref{DGPfried}) as an effective energy density component:
\be
\frac{\rho_{\rm DGP}}{3M_{\rm Pl}^2} \equiv -\frac{H}{r_c}\,,
\ee
where we have chosen the normal branch. Figure~\ref{DGPcomp} compares the evolution of $|\rho_{\rm DGP}|$ in DGP cosmology with the galileon energy density
$|\rho_\pi|$ in our galileon theory, in each case with $r_c =15$~Gpc. For simplicity, we fix the matter, radiation
and cosmological constant contributions. We see that $\rho_{\rm DGP}$ and $\rho_\pi$ agree remarkably well for the entire
expansion history until $z=0$, and begin to diverge in the future ($z<0$) when $|\pi| \sim M_{\rm Pl}$.

\section{Growth of Density Perturbations}
\label{pert}

By virtue of its non-minimal coupling to gravity, the galileon enhances the gravitational attraction between particles. 
Translated to the cosmological context, we expect more rapid growth of density perturbations. Furthermore, from the discussion of
Sec.~\ref{decouple}, the galileon enhancement should be suppressed at early times, due to self-screening,
but should become important once the matter density has dropped sufficiently.

Since the matter action is independent of the galileon in Jordan frame, the evolution of matter density perturbations, $\delta_{\rm m} \equiv \delta\rho_{\rm m}/\rho_{\rm m}$, on sub-Hubble scales is governed by the standard expression
\be
\ddot{\delta}_{\rm m} + 2H\dot{\delta}_{\rm m} = \frac{\vec{\nabla}^2\Phi}{a^2}\,,
\ee
where $\Phi$ is the Newtonian potential in Jordan frame. The effects of $\pi$ are all encoded in its contribution to the Poisson equation.

Consider expanding the galileon around its cosmological profile, $\pi(\vec{x},t) = \bar{\pi}(t) + \varphi(\vec{x},t)$.
In the Newtonian approximation, we are justified in neglecting time derivatives of $\varphi$ relative
to spatial gradients: $\vert\dot{\varphi}\vert\,\ll\,\vert\vec{\nabla}\varphi\vert$. We can thus expand~(\ref{pieomcov}) to linear order in $\varphi$ as follows,
keeping in mind that $e^{2\pi/M_{\rm Pl}}\approx 1$ for most of the expansion history, 
\be
-4r_c^2\left(\ddot{\bar{\pi}} + 2H\dot{\bar{\pi}}\right)\frac{\vec{\nabla}^2\varphi}{a^2} = M_{\rm Pl}^2\delta R\,.
\label{NR1}
\ee
Meanwhile, from the trace of~(\ref{einmod}),
\be
-M_{\rm Pl}^2\delta R \approx \delta T + 6M_{\rm Pl}\frac{\vec{\nabla}^2\varphi}{a^2}\,.
\label{NR2}
\ee
Note that we have neglected the $\pi$-dependent terms in the second line of~(\ref{einmod}), since the backreaction of the galileon
is negligible for all times, except in the recent past. Since $\delta T = - \rho_{\rm m} \delta_{\rm m} $ for non-relativistic sources, combining~(\ref{NR1}) and~(\ref{NR2}) gives
\be
\frac{\vec{\nabla}^2\varphi}{M_{\rm Pl}} = \frac{4\pi G}{3\beta}a^2\rho\;\delta_{\rm m}\,,
\label{galpoi}
\ee
where 
\be
\beta \equiv 1-\frac{2r_c^2}{3M_{\rm Pl}}\left(\ddot{\bar{\pi}} + 2H\dot{\bar{\pi}}\right)\,.
\label{betaus}
\ee
Moreover, using~(\ref{galpoi}), it is straightforward to obtain the modified evolution equation for density perturbations:
\be
\ddot{\delta}_{\rm m} + 2H\dot{\delta}_{\rm m} = 4\pi G\rho_{\rm m}\delta_{\rm m}\left(1+ \frac{1}{3\beta}\right)\,.
\label{delus}
\ee
This Poisson equation for $\varphi$ exhibits the expected features: at early times, when $r_c^2|\ddot{\bar{\pi}} + 2H\dot{\bar{\pi}}|/M_{\rm Pl}\gg 1$, the coupling to $\delta_{\rm m}$ is much suppressed since $\beta\rightarrow\infty$. This is the cosmological self-screening mechanism, which decouples
the galileon from matter inhomogeneities. At late times, on the other hand, $\beta$ becomes of order unity, and the galileon couples to matter with strength comparable
to that of gravity. The evolution equation~(\ref{delus}) takes on a standard form in the late-time regime, except for the fact that the gravitational attraction is enhanced by 
the galileon factor.

\begin{figure}[h!] %  figure placement: here, top, bottom, or page
\begin{center}
\includegraphics[width=0.9\textwidth]{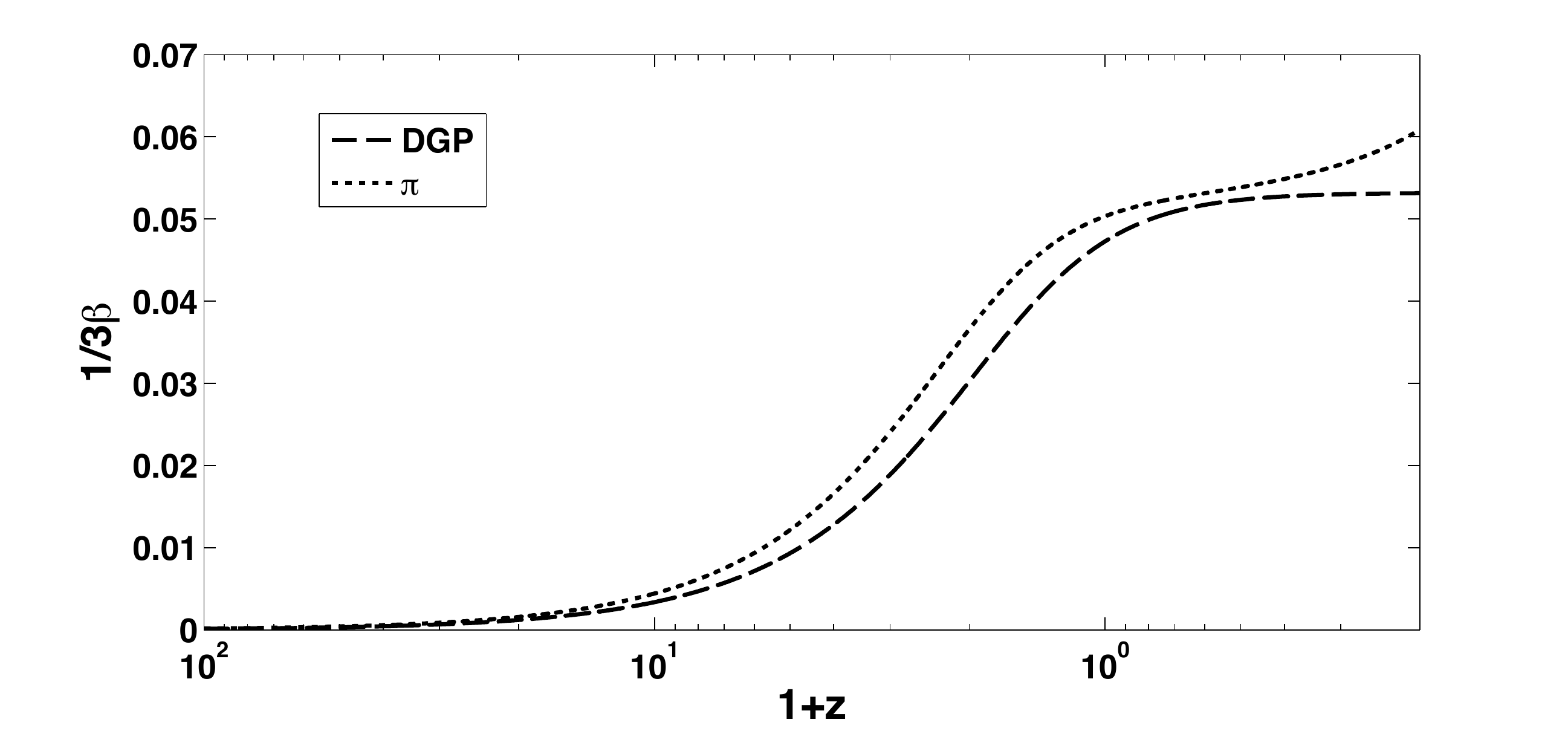}
\end{center}
   \caption{Comparison of the perturbation enhancement factor $1/3\beta$ in our galileon model (dotted) with the corresponding factor $1/3\beta_{\rm DGP}$ in DGP cosmology (dashed). As in Fig.~\ref{DGPcomp}, we have used $r_c = 15$~Gpc in each case and kept the matter, radiation and cosmological constant densities fixed.}
   \label{DGPcomppert}
\end{figure}

For comparison, the linearized perturbation equation derived for the full DGP model, derived in~\cite{lue} for spherical top-hat perturbations, takes
on an identical form to~(\ref{delus}), with $\beta$ in this case given by
\be
\beta^{\rm DGP} = \frac{1+2r_cH+r_c^2H^2}{1+r_cH}\,.
\label{betaDGP}
\ee
(Note that the choice of sign is consistent with the normal branch of DGP.) Clearly the asymptotic behavior of $\beta^{\rm DGP}$ for $Hr_c\ll 1$ and $Hr_c\gg 1$
agrees with that of~(\ref{betaus}). Figure~\ref{DGPcomppert} compares $1/3\beta$ in our model with $1/3\beta_{\rm DGP}$ in DGP cosmology, by solving numerically for the respective cosmological backgrounds. As in the discussion of Sec.~\ref{DGPc}, we take $r_c = 15$~Gpc in each case and fix the energy density in matter, radiation and cosmological term. Given the close agreement shown in the figure, we expect our galileon theory to make nearly identical predictions for structure formation as the full DGP model. We leave for future work a detailed study of the evolution of perturbations in the covariant galileon theory.

\section{Observational Constraints}
\label{obssec}

A rigorous comparison of the galileon cosmological scenario with observations requires a full likelihood analysis, which is beyond the scope of this paper. For the purpose of this Section, we restrict ourselves to a comparison with a few observables and derive a lower bound on $r_c$. Since $r_c$ is the only parameter that we allow to vary, our estimate is likely conservative.
\subsection{Mass estimates}

Because of the non-minimal coupling of the galileon, the fractional matter density, $\Omega_{\rm m}$, depends on $\pi$, as seen from~(\ref{Omi}):
\be
\Omega_{\rm m} = \frac{\rho_{\rm m}e^{2\pi/M_{\rm Pl}}}{3H^2M_{\rm Pl}^2}\,.
\ee 
Therefore, for fixed matter density today, $\Omega_{\rm m}^{(0)}$, the matter density in the past will differ from the standard gravity prediction by
\be
\left.\frac{\Omega_{\rm m}}{\Omega_{\rm m}^{\rm std}}\right\vert_{z\;\gsim\; 1} = e^{-2\pi_0/M_{\rm Pl}}\,,
\ee
where $\pi_0$ is the present value of the galileon. And since $\pi_0< 0$ on the stable branch, $\Omega_{\rm m}$ is {\it larger} at early times than predicted by standard gravity, again keeping $\Omega_{\rm m}^{(0)}$ fixed. Hence $\pi_0$ is constrained by estimates of the matter density at various redshifts, such as from cluster counts, Lyman-$\alpha$ forest and weak lensing observations. (Similar considerations apply to coupled dark matter-dark energy models~\cite{piersteandme}.)

\begin{figure}[h!] %  figure placement: here, top, bottom, or page
\begin{center}
\includegraphics[width=0.75\textwidth]{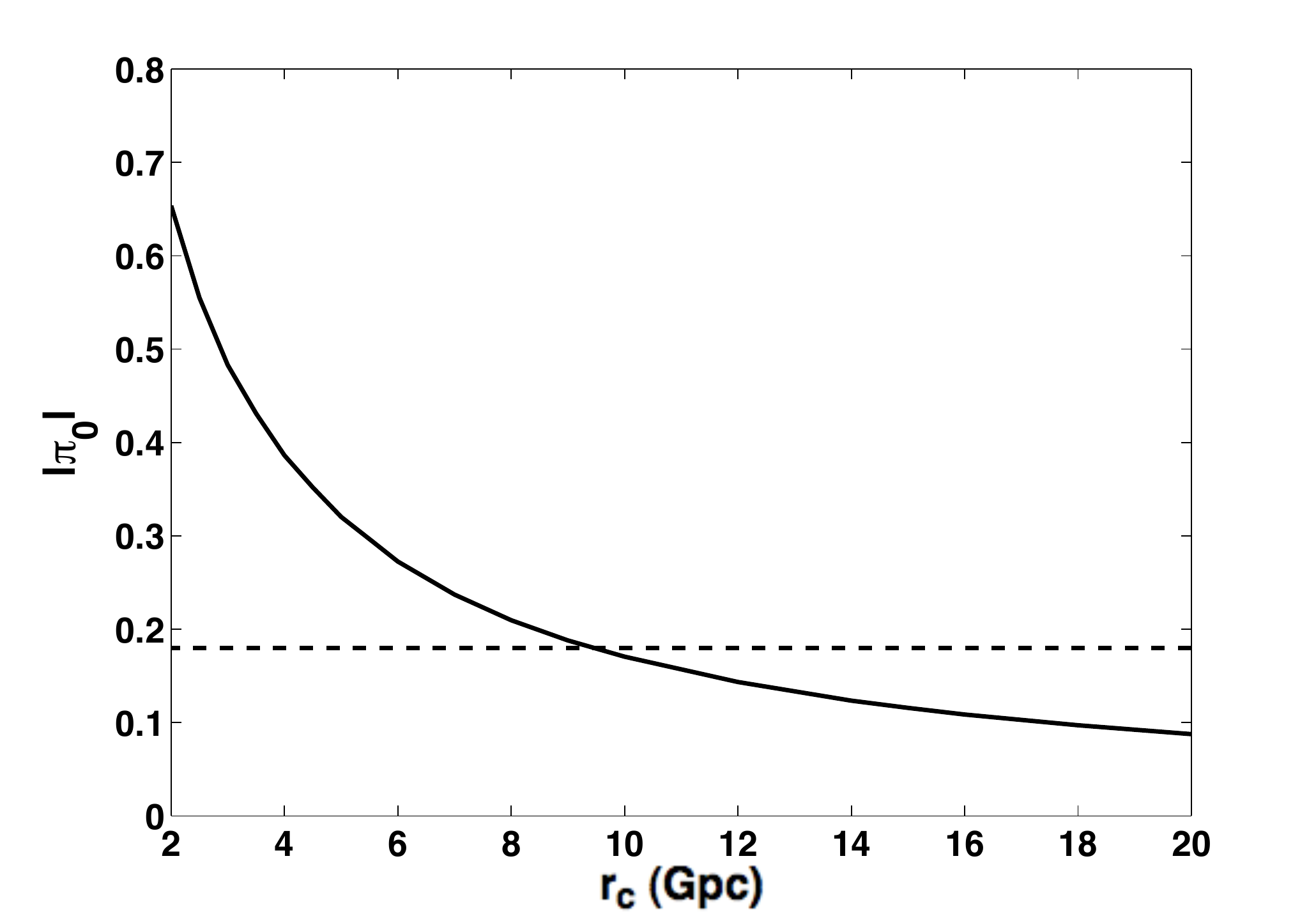}
\end{center}
\caption{Value of the galileon field at the present time as a function of $r_c$. For each value of $r_c$, the cosmological term is adjusted to keep
$\Omega_{\rm m}^{(0)}= 0.26$ fixed. The dashed line represents the threshold value of $0.18\;M_{\rm Pl}$ from mass estimates --- see 
the bound in~(\ref{rcbound}).}
\label{pi0}
\end{figure}

Although the constraints from these observables should be revisited in the presence of a galileon, it has been argued that the allowed
range of $\Omega_{\rm m}^{(0)}$ is generally insensitive to the specifics of dark energy~\cite{pierste}. A general analysis combining SNIa Gold data set~\cite{gold}, Wilkinson Anisotropy Microwave Probe (WMAP) power spectra~\cite{wmap1}, and Two-Degree Field (2dF) galaxy survey~\cite{2df} obtained $0.23\; \lsim\; \Omega_{\rm m}^{(0)} \;\lsim\; 0.33$~\cite{pierste}. (See also~\cite{other1,other2}.) In the absence of a full likelihood analysis, we can obtain a conservative bound on $\pi_0$ by requiring that $e^{-2\pi_0/M_{\rm Pl}} \; \lsim \; 0.33/0.23$, or 
\be
|\pi_0| \; \lsim \; 0.18\; M_{\rm Pl}\,. 
\label{rcbound}
\ee
Figure~\ref{pi0} shows $|\pi_0|$ for various values of $r_c$, as obtained numerically. We can read off that the above bound on $|\pi_0|$ is satisfied for
$r_c \;\gsim \; 10\; {\rm Gpc}$, or roughly 3 times larger than the radius of the observable universe. We will see shortly that luminosity
distance observations tighten the bound to $r_c\;\gsim \; 15$~Gpc.

\subsection{Cosmological Distances}

\begin{figure}[h!] %  figure placement: here, top, bottom, or page
\begin{center}
\includegraphics[width=0.75\textwidth]{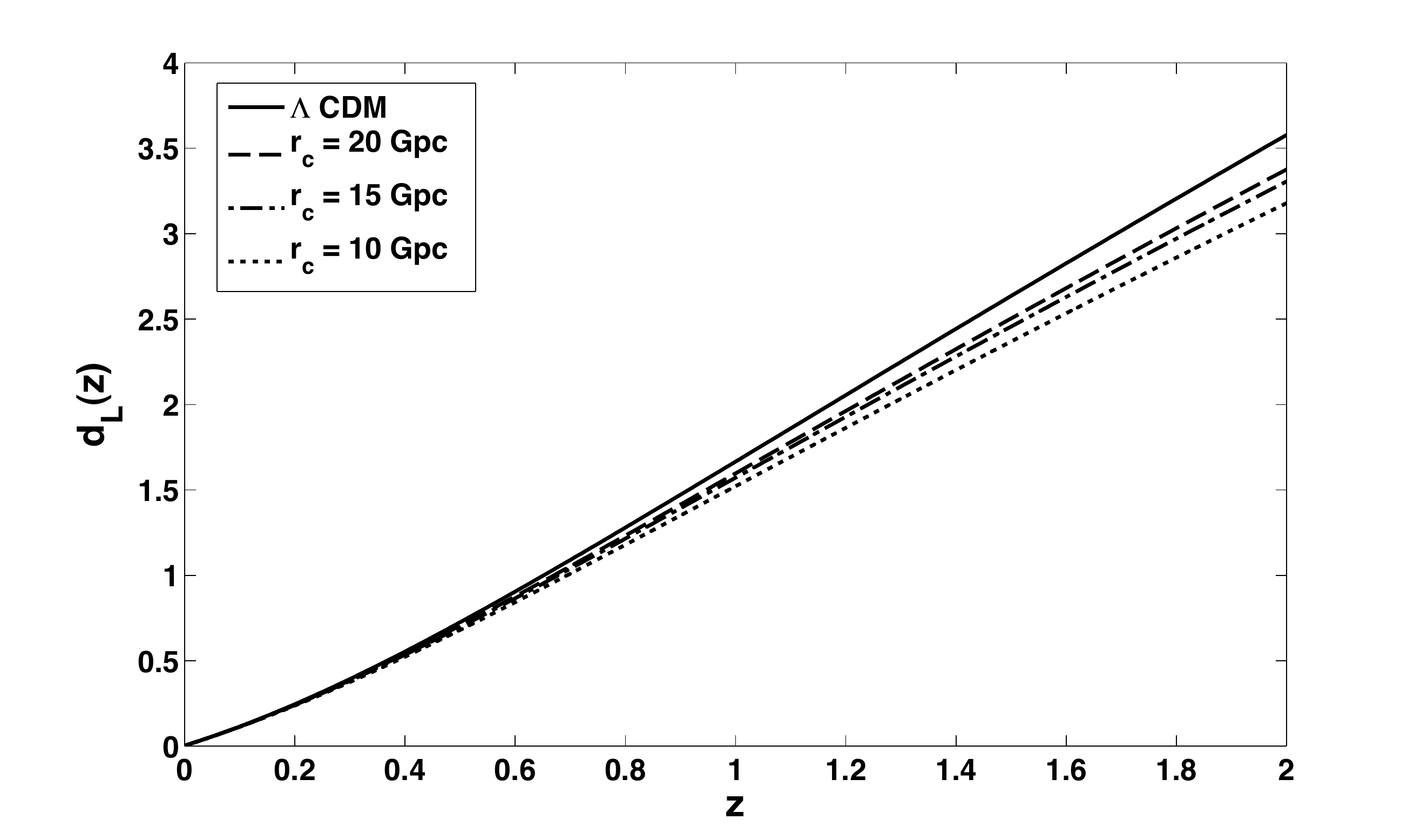}
\end{center}
   \caption{Luminosity distance ($d_{\rm L}$) as a function of redshift for $\Lambda$CDM (solid), and our galileon model with $r_c = 10$~Gpc (dotted), $r_c = 15$~Gpc (dash-dotted) and $r_c = 20$~Gpc (dashed). In each case, we have fixed $\Omega_{\rm m}^{(0)} = 0.26$. } 
   \label{dL}
\end{figure}

Next we turn our attention to cosmological distance tests, in particular the luminosity distance relation, 
\be
d_{\rm L}(z) =(1+z)\int_0^z \frac{{\rm d}z'}{H(z')}\,,
\ee
as constrained by Type Ia supernovae (SNIa). Figure~\ref{dL} shows the luminosity distance relation with $\Omega_{\rm m}^{(0)} = 0.26$ for the $\Lambda$CDM model, and for our galileon model with $r_c = 10$, $15$ and $20$~Gpc. In the galileon examples, $\Lambda$ was adjusted in order to keep $\Omega_{\rm m}^{(0)}$ fixed. 

\begin{figure}[h!] %  figure placement: here, top, bottom, or page
\begin{center}
\includegraphics[width=0.75\textwidth]{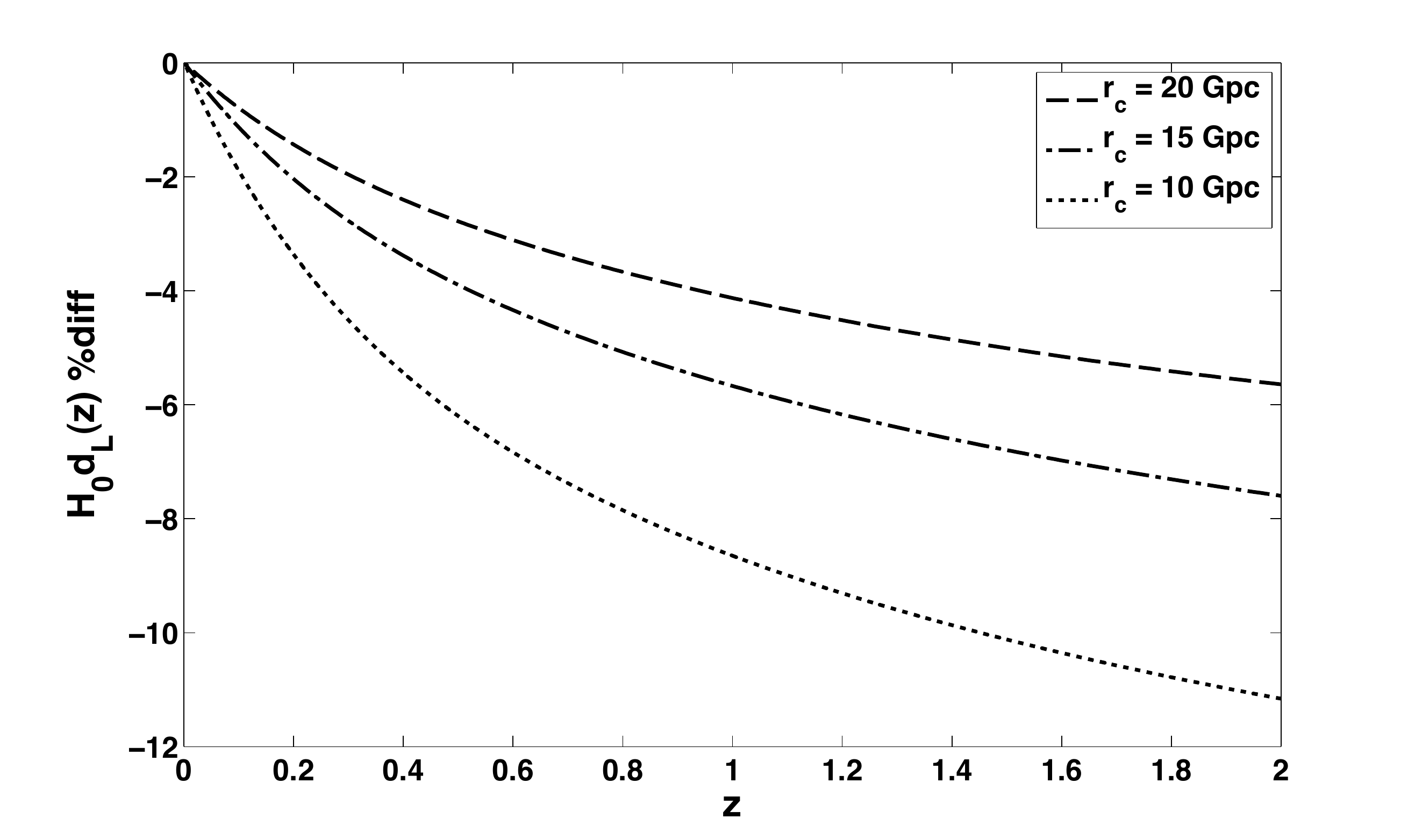}
\end{center}
   \caption{Percentage difference between the galileon examples and $\Lambda$CDM.} 
   \label{dL2}
\end{figure}

Figure~\ref{dL2} shows the percentage difference between the various galileon examples and the $\Lambda$CDM fiducial model. The uncertainties in present SNIa data constrain the luminosity distance to no better than $\sim 7\%$ over the range $0< z < 1.5$. Therefore, from the percentage differences shown in the figure, we see that  luminosity-distance observations tighten the bound on $r_c$ to
\be
r_c \;\gsim \; 15\; {\rm Mpc}\,.
\label{rcbound2}
\ee
This constitutes our main constraint on the scale of the modification.

Another distance constraint comes from the angular-diameter distance to the last scattering surface, 
\be
d_{\rm A}(z_{\rm rec}) = \frac{1}{1+z_{\rm rec}} \int_0^{z_{\rm rec}} \frac{{\rm d}z'}{H(z')}\,,
\ee
which determines the position of the cosmic microwave background (CMB) acoustic peaks. For $r_c\;\gsim\; 15$~Gpc, the difference in $d_{\rm A}(z_{\rm rec})$ compared to $\Lambda$CDM is less than 10\%, again keeping $\Omega_{\rm m}^{(0)}$ fixed, which is within current CMB uncertainties. Note that a similar constraint $\l r_c\;\gsim\; 3-3.5H_0^{-1}\r$ was obtained recently by confronting the DGP normal-branch cosmology against CMB, SNIa and Hubble constant observations~\cite{huseljak}.

\section{Conclusions}
\label{conclusions}

In this paper we have studied the cosmology of a galileon field theory, obtained by covariantizing the $\pi$-Lagrangian of the DGP model. Despite
being a local theory in 3+1 dimensions, the resulting cosmological evolution is remarkably similar to that of the full 4+1-dimensional DGP framework,
at least for $Hr_c \;\gsim\; 1$ (or $|\pi|\;\lsim\; M_{\rm Pl}$). The similarity holds for both the expansion history (Fig.~\ref{DGPcomp}) and the evolution
of density perturbations (Fig.~\ref{DGPcomppert}).

In particular, as in the DGP model our covariant galileon theory yields two branches of solutions, depending on the sign of $\dot{\pi}$.
Perturbations are stable on one branch and ghost-like on the other. The effective equation of state for the galileon is phantom-like ($w_\pi < -1$)
on the stable branch, and standard ($w_\pi>-1$) on the unstable branch. A key difference, however, is that the galileon field theory does not generate
a self-accelerated solution --- as shown in Sec.~\ref{selfac}, the would-be self-accelerated solution with $H\sim 1/r_c$ is spoiled by $1/Hr_c$
corrections in our theory.

An interesting effect uncovered in our analysis is a cosmological version of the self-screening (or Vainshtein) mechanism. At early times, $Hr_c\gg 1$,
the evolution of $\pi$ is dominated by the self-interactions terms. In turn, this results in $\pi$ being a negligible component, with its energy density suppressed
by a factor ${\cal O}(1/Hr_c)$ compared to matter and radiation. This cosmological self-screening is crucial in the recovery of standard cosmology at early times.

Once the expansion rate drops to $Hr_c \sim 1$, however, the galileon becomes an important player in the Friedmann equation. A preliminary analysis
of observational constraints in Sec.~\ref{obssec} shows that the modifications to the expansion history are consistent with the observed late-time cosmology
provided $r_c\;\gsim\; 15$~Gpc. We should emphasize that this bound is most certainly conservative and will be relaxed by allowing other cosmological parameters to vary.

Our analysis offers a host of interesting avenues to explore:

\begin{itemize}

\item Following up on the discussion of the last paragraph, a thorough comparison with observations requires a full likelihood analysis, allowing
various cosmological parameters (such as $\Omega_{\rm m}^{(0)}$, $h$, etc.) to vary.

\item A study of the implications for structure formation requires a more rigorous treatment of inhomogeneities in the presence of the
galileon. To probe the non-linear regime, it should be straightforward to generalize the N-body simulations of~\cite{nbody1, nbody2} to include the galileon
field theory studied here.

\item In this work we have focused exclusively on the cubic interaction term that arises in the decoupling limit of DGP. Recently,~\cite{galileon} derived the most general galileon field theory (without gravity) which is invariant under the Galilean shift symmetry $\partial_\mu\pi\rightarrow\partial_\mu\pi + c_\mu$, and whose equations of motion are second order. It was shown in~\cite{deff} that the equations remain of second order in the presence of gravity, provided $\pi$ is suitably coupled to gravity. It would be very interesting to extend our analysis to include these higher-order interactions terms.

\item The extension of our 4$D$ theory to include multiple galileon fields should offer a reliable proxy
for the cosmology of higher-dimensional DGP models, such as Cascading Gravity~\cite{cascade1,cascade2,claudiareview}. 
In this construction, our 3-brane lies within a succession of higher-dimensional DGP branes,
embedded in one another within a flat bulk space-time. The corresponding 4$D$ covariant theory
should therefore include 2 interacting galileon fields, each with its own $r_c$ scale.

\end{itemize}

{\bf Acknowledgments}
We thank Pier-Stefano~Corasaniti, Michel~Gingras, James~Taylor, Brian McNamara, Rob~Myers, Alberto~Nicolis, Roman~Scoccimarro, Andrew~Tolley, Mark~Trodden, and especially Mark~Wyman for helpful discussions, as well as Wayne~Hu and Kazuya~Koyama for initial discussions. This work was
supported in part at the Perimeter Institute by the Government of Canada through NSERC and by the Province of Ontario through the Ministry of Research \& Innovation.\\

\end{document}